\documentclass[fleqn,usenatbib]{mnras}

\usepackage{newtxtext,newtxmath}

\usepackage[T1]{fontenc}
\usepackage{ae,aecompl}

\usepackage{adjustbox}
\usepackage{pifont}
\usepackage{float}
\usepackage{graphicx}	
\usepackage{amsmath}	
\usepackage{multirow}
\usepackage{caption}
\usepackage{subcaption}
\usepackage{gensymb}
\usepackage{booktabs}
\usepackage{times}
\usepackage{xcolor}
\usepackage{ulem}
\usepackage{hyperref}

\usepackage{etoolbox}

\title[Dust Traps and the Formation of Cavities in TDs]{Dust Traps and the Formation of Cavities in Transition Discs: A millimetre to sub-millimetre comparison survey}

\author[Norfolk et al.]{
Brodie J. Norfolk$^{1}$\thanks{E-mail: bnorfolk@swin.edu.au (SUT)}, Sarah T. Maddison$^{1}$, Christophe Pinte$^{2,3}$, \newauthor
Nienke van der Marel$^{4,5}$,  Richard A. Booth$^{6}$, Logan Francis$^{4,5}$, Jean-Fran\c{c}ois Gonzalez$^{7}$, \newauthor
Fran\c{c}ois M\'enard$^{3}$, Chris M. Wright$^{8}$, Gerrit van der Plas$^{3}$, Himanshi Garg$^{2}$ 
\\
$^{1}$Centre for Astrophysics and Supercomputing (CAS), Swinburne University of Technology, Hawthorn, Victoria 3122, Australia\\
$^{2}$Monash Centre for Astrophysics (MoCA) and School of Physics and Astronomy, Monash University, Clayton Vic 3800, Australia\\
$^{3}$Univ. Grenoble Alpes, CNRS, IPAG, F-38000 Grenoble, France\\
$^{4}$Department of Physics \& Astronomy, University of Victoria, Victoria, BC V8P 5C2\\
$^{5}$Herzberg Astronomy \& Astrophysics Programs, National Research Council of Canada, 5071 West Saanich Road, Victoria BC V9E 2E7, Canada\\
$^{6}$Institute of Astronomy, University of Cambridge, Madingley Road, Cambridge, CB3 0HA, UK\\
$^{7}$Univ Lyon, Univ Claude Bernard Lyon 1, ENS de Lyon, CNRS, Centre de Recherche Astrophysique de Lyon UMR5574, F-69230, Saint-Genis-Laval, France \\
$^{8}$School of Science, University of New South Wales, PO Box 7916, Canberra, BC 2610, Australia
}

\date{Accepted XXX. Received YYY; in original form ZZZ}

\pubyear{2015}

\begin{document}
\label{firstpage}
\pagerange{\pageref{firstpage}--\pageref{lastpage}}
\maketitle

\begin{abstract}
The origin of the inner dust cavities observed in transition discs remains unknown. The segregation of dust and size of the cavity is expected to vary depending on which clearing mechanism dominates grain evolution. We present the results from the Discs Down Under program, an 8.8~mm continuum Australia Telescope Compact Array (ATCA) survey targeting 15 transition discs with large ($\gtrsim$ 20\,au) cavities, and compare the resulting dust emission to Atacama Large millimetre/sub-millimetre Array (ALMA) observations. Our ATCA observations resolve the inner cavity for 8 of the 14 detected discs. We fit the visibilities and reconstruct 1D radial brightness models for 10 sources with a S/N > $5\sigma$. We find that, for sources with a resolved cavity in both wavebands, the 8.8\,mm and sub-mm brightness distributions peak at the same radius from the star. We suggest that a similar cavity size for 8.8\,mm and sub-mm dust grains is due to a dust trap induced by the presence of a companion.
\end{abstract}




\section{Introduction}
Protoplanetary discs around young stars are the birth places of planets. Of particular interest are the so-called transition discs, which exhibit inner cavities in their dust distribution. They were historically believed to be the phase of transition between gaseous protoplanetary discs and debris discs, but their exact place in the evolutionary timeline (if any) still remains unknown \citep{2014prpl.conf..497E}. They were originally identified via a dip in the mid-infrared region of their spectral energy distributions  \citep{1989AJ.....97.1451S,1996AJ....111.2066W}, suggesting a deficit of micron-sized grains in the inner disc. Sub-millimetre interferometry has confirmed this deficit and resolved the inner cavity in the dust of numerous transition discs for the largest ($>$20 au) cavities \citep[e.g.][]{2009ApJ...704..496B,2011ApJ...732...42A,2020francis}.

The deficit of central emission may result from grain growth due to the vertical settling, radial drift, and coagulation of grains, resulting in a radial and vertical size sorting, and grain growth to sizes outside current observation capabilities \citep{2005ApJ...625..414T}. Another possible explanation is a dust trap; the trapping of dust at local maxima in the gas density \citep{1972fpp..conf..211W, 2012A&A...545A..81P} that can overcome the radial drift barrier \citep{1977MNRAS.180...57W}. Dust grains interior to  this region accrete onto the host star, depleting the inner disc and forming a cavity. A number of physical mechanisms can produce dust traps in transition discs, including photoevaporation, dynamical clearing from a companion, and dead-zones \citep{2011ARA&A..49...67W, 2016A&A...596A..81P}.

As the accretion rate decreases, high-energy photons can impinge the accretion flow and photoevaporate the disc progressively further away. This limits the resupply of inner disc material and increases the local gas pressure at the photoevaporation front, trapping dust at the inner edge of the disc \citep{1994ApJ...428..654H, 2001MNRAS.328..485C, 2006MNRAS.369..229A}. Numerous photoevaporation models \citep{2012MNRAS.426L..96O, 2015ApJ...804...29G, 2017MNRAS.472.2955O, 2017RSOS....470114E, 2018MNRAS.473L..64E} suggest that discs with an inner deficit of emission due to photoevaporative winds will have low accretion rates (\(\rm \leq10^{-9} ~M_{\odot} yr^{-1}\)) and exhibit cavities with small radii (\(\rm \leq 10~au\), see Figure 5 in \citet{2012MNRAS.426L..96O}, or \(\rm \leq 20~au\), see Figure 6 in \citet{2017RSOS....470114E}). This is inconsistent with (sub)millimetre observations of transition discs, exhibiting cavities of tens of au in size and high accretion rates \citep{2014A&A...568A..18M}.

Dust traps can also result from the dynamical clearing of a companion. The formation of a planetary mass companion will locally deplete gas and/or dust in the disc and produce a pressure bump at the cavity edge \citep[e.g.][]{2004A&A...425L...9P,2006MNRAS.373.1619R, 2007A&A...474.1037F, 2010A&A...518A..16F, 2012A&A...545A..81P}, while the tidal truncation of the disc by a binary companion \citep[e.g.][]{1979MNRAS.188..191L} results in a gas pressure bump at the leading edge of the truncation, consequentially trapping dust and forming a cavity as the inner disc accretes onto the host-star at viscous timescales (\(\leqslant 10^5\) years) \citep{2007A&A...469.1169B}. Dust traps formed by the presence of a companion are expected to result in a grain size distribution that mirrors the profile of the gas pressure at the maxima with larger grains strongly concentrating at the peak of the pressure profile, and smaller grain exhibiting a more extended radial structure. This is due to the relationship between the grains Stokes number and its size, as larger grains with \(\rm St \sim 1\) are significantly influenced by the pressure of the gas whereas smaller grains with a \(\rm St < 1\) are more coupled to the gas. Disc cavities have been attributed to the presence of a companion in a number of previous studies \citep{2016A&A...585A..58V, 2017A&A...607A..55V, 2018ApJ...854..177V, 2018ApJ...859...32P, 2018A&A...616A..79G, 2018MNRAS.477.1270P, 2020ApJ...888L...4T}.

In regions of low ionization where the magneto-rotational instability is inhibited, there is an increase of accretion stress and subsequent local maxima in the gas pressure that can trap dust \citep{2015A&A...574A..68F, 2016A&A...590A..17R}. Models based on dust trapping at dead zone edges \citep{2016A&A...596A..81P, 2019ApJ...871...10U} are highly dependent on the level of turbulence throughout the disc, and predict a similar peak in the radial distribution of sub-mm to mm dust species and strong gas depletion in the outer disc. 

More than one of these physical mechanisms may occur simultaneously in transition discs. However, the dominant physical mechanism that produces a gas pressure maxima (dust trap) will result in distinct cavity sizes, accretion rates, and/or grain size distributions. As different wavebands trace dust grains of order the same size as the wavelength, multi-wavelength observations of transition discs can reveal differences in the grain size distribution and further differentiate between the mechanisms behind dust trap and cavity formation. 

Previous multi-wavelength observations of transition discs that probe both sub-millimetre and millimetre sized grains \citep{2015MNRAS.453..414W,2015ApJ...812..126C,2018ApJ...865...37M} show that each population is trapped at a similar radial position from the host star \citep{2015ApJ...810L...7V}. However, the number of sources with multi-wavelength data still remains small. In this paper we present new millimetre observations for 15 transition discs combined with archival ALMA millimetre observations, which is the largest multi-wavelength survey of transition discs to date.

\section{Observations and Data Reduction} \label{sect:3}
\subsection{Sample Selection}

Recent high-resolution (\(\leqslant0.2^{\prime\prime}\)) observations of transition discs with ALMA have provided a growing catalogue of discs observed at sub-mm wavelengths which show resolved dust cavities. Our Discs Down Under survey was designed to specifically complement this catalogue at 8.8~mm for southern hemisphere sources that are observable by the ATCA.

Our sample (see Table \ref{tab:1}) includes four Herbig Ae stars (HD34282, HD100453, HD135344B, HD169142), and eleven T Tauri stars (SZ~Cha, CS~Cha, HP~Cha, HD143006, RY~Lup, J1604, J160830.7, Sz111, SR24S, SR21, DoAr44). Three of these are located in Lupus (RY~Lup, J16083070, \& Sz111), three in Chamaeleon (SZ~Cha, CS~Cha, \& HP~Cha) and three in Ophiuchus (SR~24S, SR~21 and DoAr44). One target (J1604) is a member of the Upper Scorpius Association. Three of the targets have previously been observed with ATCA in the 7~mm band at low resolution and were unresolved (Sz111 and RY~Lup, \citealt{2010A&A...515A..77L}; CS~Cha, \citealt{2009A&A...495..869L}), and one with the VLA at 7~mm \citep[HD169142,][]{2006MNRAS.365.1283D}.

This survey also targeted the protoplanetary disc HD163296, which is not a transition disc and hence is excluded from the analysis. Observational results for this source are presented in Appendix \ref{app:hd163296}.

\begin{table*}
    \centering
    \caption{Discs Down Under survey source list. Distances for our sample are taken from \citet{2018yCat.1345....0G}, except for HP~Cha which is taken from \citet{1997A&A...327.1194W}. We extract the mass accretion rates of SZ~Cha from \citet{2019A&A...631L...2M}, HD143006 from \citet{2020A&A...639A..58M}, J1608 and Sz111 from \citet{2017A&A...600A..20A}, and the remaining sources from \citet{2020francis}.}
    \label{tab:1}
    \begin{adjustbox}{width=0.99\textwidth}
        \begin{tabular}{lcccccc}
        \hline
        Target                  & RA (ICRS)     & Dec. (ICRS)    & Distance (pc)  & ATCA Observing Date & ALMA ID & $\dot{M}$ (\(\rm log(M_{\odot}/yr)\)) \\ \hline
        HD 34282            & 05 16 00.48 & -09 48 35.4    & 325  & 10/06/2018 &  2015.1.00192.S, 2017.1.01578.S &  \(\rm <-8.30\)\\
        SZ Cha               & 10 58 16.75 & -77 17 17.2   & 189  & 03/05/2016 &  2013.1.00437.S & \(\rm -7.65\)\\
        CS Cha               & 11 02 24.89 & -77 33 35.7   & 176  & 29/04/2016 & 2017.1.00969.S &  \(\rm  -8.30\)\\
        HP Cha               & 11 08 15.09  & -77 33 53.2    & 160 & 22/05/2017 & 2017.1.01460.S & \(\rm -8.97\)\\
        HD 100453               & 11 33 05.58 & -54 19 28.5   & 104  & 04/05/2016 & 2015.1.00192.S, 2017.1.01424.S & \(\rm <-8.30\)\\
        HD 135344B              & 15 15 48.45 & -37 09 16.0    & 135  & 13/05/2016 & 2017.1.00884.S &  \(\rm -7.37\)\\
        HD 143006               & 15 58 36.91 & -22 57 15.2   & 166 & 04/05/2017 &  2016.1.00484.L & \(\rm -7.79\)\\
        RY Lup               & 15 59 28.39 & -40 21 51.3 & 159  & 06/05/2016 \& 06/05/2017 & 2017.1.00449.S &  \(\rm -8.20\)\\
        J1604         & 16 04 21.66   & -21 30 28.4  & 145 & 23/04/2016 \& 12/05/2017 & 2015.1.00888.S & \(\rm -10.54\)\\
        J1608  & 16 08 30.69 & -38 28 26.9      & 156 & 22/04/2016 \& 13/05/2017 & 2012.1.00761.S & \(\rm -9.1\)\\
        Sz111               & 16 08 54.68 & -39 37 43.2 & 158 & 07/05/2016 \& 06/05/2017 & 2012.1.00761.S & \(\rm -9.1\)\\
        SR24S               & 16 26 58.50 & -24 45 36.7   & 114 & 17/05/2016 & 2017.1.00884.S &  \(\rm -7.15\)\\
        SR21               & 16 27 10.28 & -24 19 12.6     & 138 & 25/04/2016 & 2017.1.00884.S & \(\rm -7.90\)\\
        DoAr44               & 16 31 33.46 & -24 27 37.2  & 120 & 24/04/2016 & 2012.1.00158.S & \(\rm -8.20\)\\
        HD 169142               & 18 24 29.78 & -29 46 49.3  & 117 & 15/05/2016 & 2016.1.00344.S & \(\rm -8.70\)\\ \hline
\end{tabular}
    \end{adjustbox}
\end{table*}

\subsection{ATCA} \label{atca_reduction}
We used the ATCA radio telescope at 34~GHz (8.8~mm) to conduct our survey from 2016 to 2018 (project code C3119). The Compact Array Broadband Backend (CABB) \citep{2011MNRAS.416..832W} provides observations with two bands that contain \(\rm 2048 \times 1\)~MHz channels, which we centred at 33~GHz and 35~GHz. Observations were conducted in the 6A array configuration which has baseline lengths ranging from 337~m to 5939~m and a theoretical resolution of $0.3^{\prime\prime}$. The synthesised beam for each observation is detailed in Table~\ref{tab:3}. The astrometric accuracy of ATCA observations is primarily determined by the atmospheric conditions and the properties of the phase calibrators. The weather during the observations varied for each science target, and the seeing monitor RMS path length noise for each observation is summarised in Table \ref{tab:atca}.

The science targets were observed with {a sequence of 10~min on-source integration and 2~min} integration of the gain/phase calibrator. The bandpass and flux calibrators were observed for \(\sim\)15~min and pointing checks were made on the phase calibrator every \(\sim\)60-90~min. All observational and calibration details are summarised in Table~\ref{tab:atca}.

\begin{table}
\caption{ATCA observing log.}
\label{tab:atca}
\centering
\begin{adjustbox}{width=0.48\textwidth}
\begin{tabular}{cccccc}
\hline
\multirow{2}{*}{Date} & \multirow{2}{*}{\begin{tabular}[c]{@{}c@{}}Obs. \\ Time\\ (min)\end{tabular}} & \multicolumn{3}{c}{Calibrators} & \multirow{2}{*}{\begin{tabular}[c]{@{}c@{}}Ave. \\ seeing\\ rms (\(\rm \mu m \))\end{tabular}} \\ \cline{3-5} &  & Band-pass & Flux & \begin{tabular}[c]{@{}c@{}}Gain/\\ Phase\end{tabular} \\  \hline
22/04/16 & 225 & 1253-055 & 1934-638 & 1606-39 & 380 \\
23/04/16 & 225 & 1253-055 & 1934-638 & 1622-253 & 388\\
24/04/16 & 430 & 1253-055 & 1934-638 & 1622-253 & 248\\
25/04/16 & 450 & 1253-055 & 1934-638 & 1622-253 & 152\\
29/04/16 & 430 & 0537-441 & 1934-638 & 1057-797 & 202\\
03/05/16 & 340 & 0537-441 & 1934-638 & 1057-797 & 308\\
04/05/16 & 450 & 0537-441 & 1934-638 & 1129-58 & 95\\
06/05/16 & 450 & 0537-441 & 1934-638 & 1606-39 & 155\\
07/05/16 & 450 & 1253-055 & 1934-638 & 1606-39 & 170\\
13/05/16 & 450 & 1253-055 & 1934-638 & 1451-400 & 99\\
15/05/16 & 450 & 1253-055 & 1934-638 & 1804-251 & 235\\
17/05/16 & 340 & 1253-055 & 1934-638 & 1622-310 & 300\\
04/05/17 & 430 & 1253-055 & 1934-638 & 1622-253 & 185\\
05/05/17 & 450 & 1253-055 & 1934-638 & 1606-39 & 114\\
06/05/17 & 430 & 1253-055 & 1934-638 & 1606-39 & 139\\
12/05/17 & 340 & 1253-055 & 1934-638 & 1622-253 & 236\\
13/05/17 & 430 & 1253-055 & 1934-638 & 1606-39 & 196\\
22/05/17 & 320 & 0537-441 & 1934-638 & 1057-797 & 313\\
10/06/18 & 450 & 1921-293 & 1934-638 & 0514-161 & 184\\
12/06/18 & 430 & 1253-055 & 1934-638 & 1817-254 & 218\\ \hline
\end{tabular}
\end{adjustbox}
\end{table}
The data was processed using the Miriad package \citep{1995ASPC...77..433S} and followed the standard procedure outlined in Section 22 of the ATCA User Guide\footnote{\url{https://www.atnf.csiro.au/computing/software/miriad/userguide/userhtml.html}}. Briefly, this involved: correcting for the frequency-dependent gain using the miriad task \textit{mfcal}; then using the flux density of the ATCA primary flux calibrator, 1937-638, to re-scale the visibilities measured by the correlator using the miriad task \textit{mfcal} with the option \textit{nopassol} set; and correcting for the gain of the system's time variable properties due to changing conditions using the miriad task \textit{gpcal}. To reduce the noise in our data while maintaining as complete an observational track as possible, we flagged all data with a seeing monitor RMS path length noise above 400~$\rm \mu$m using \textit{uvflag}, and calibrator amplitude readings that deviated more than 10\% from the mean flux using \textit{blflag}. Any unusual spikes seen in the channel vs. amplitude or the channel vs. phase plots were also flagged using \textit{uvflag}.

We extracted the ATCA fluxes by first concatenating the 33~GHz and 35~GHz observations using the \textit{uvcat} miriad task, subsequently producing observations centred at 34~GHz.  We then used the \textit{fits} miriad task to output the resulting uv data-set as a UVFITS file and binned the visibilities using python. The total integrated fluxes were taken at zero spacing (Table~\ref{tab:3}).

After calibrating the data, images at 34~GHz were produced using robust weighting of 0.5 for uniform weighting or a \textit{sup} value of 0 for natural weighting with the \textit{invert} task. The \textit{sup} parameter refers to the area around the source (in arcseconds) where \textit{invert} will attempt to suppress side-lobes. The dirty images were cleaned to $5\sigma$ (5 times the RMS noise level) using the \textit{clean} task and the beam was restored using the \textit{restor} task. The resulting images for 13 of the 14 detected sources are presented in Figure \ref{fig:all_images}. The relatively poor seeing during the 3 May 2016 synthesis track of SZ~Cha, coupled with the phase decorrelations in the visibilities, required data flagged that resulted in an unresolved image.
We therefore exclude the image of SZ~Cha and only present the flux determined during the period of best seeing in Table~\ref{tab:3}. Additionally, the 10 June 2018 synthesis track resulted in a non-detection of HD34282, likely a product of the low peak brightness due to its distance. As a result we only present the 3$\sigma$ upper limit in Table~\ref{tab:3}.

\begin{table*}
    \centering
    \caption{Discs Down Under survey results of ATCA and ALMA continuum observations.}
    \label{tab:3}
    \begin{adjustbox}{width=1\textwidth}
        \begin{tabular}{lcccccccc}
        \hline
        \multirow{2}{*}{Target} & \multirow{2}{*}{\begin{tabular}[c]{@{}c@{}}ATCA \\ 34~GHz flux \\ (mJy)\end{tabular}} &
        \multirow{2}{*}{\begin{tabular}[c]{@{}c@{}} ATCA \\ \(1\sigma\) rms \\  (mJy)\end{tabular}} & \multirow{2}{*}{\begin{tabular}[c]{@{}c@{}} ATCA \\ \(\theta_{\rm beam}\) \\ ($\rm as \times as$)\end{tabular}} &  
        \multirow{2}{*}{\begin{tabular}[c]{@{}c@{}}ALMA \\ freq. \\ (GHz)\end{tabular}} &
        \multirow{2}{*}{\begin{tabular}[c]{@{}c@{}}ALMA \\ flux \\  (mJy)\end{tabular}} &
        \multirow{2}{*}{\begin{tabular}[c]{@{}c@{}} ALMA\\ \(1\sigma\) rms \\  (mJy)\end{tabular}} & 
        \multirow{2}{*}{\begin{tabular}[c]{@{}c@{}} ALMA \\ \(\theta_{\rm beam}\) \\ ($\rm as \times as$)\end{tabular}} &
        \multirow{2}{*}{\begin{tabular}[c]{@{}c@{}}Spec. \\ Slope \\ (\(\rm \alpha_{mm}\))\end{tabular}} \\ \\ \\ \midrule
        HD34282            & <0.290\(^{a}\) & 0.125 & 3.0 x 0.38 & 255 & 99 & 0.10 & 0.06 x 0.05 & 2.74\\
        SZ Cha               & 0.227 & 0.036 & 0.58 x 0.32 & 336 & 294 &  1.65  & 0.21 x 0.12 & 3.64\\
        CS Cha               & 0.483 & 0.062  & 0.43 x 0.39 & 341 & 165 &  0.10 & 0.07 x 0.04 & 2.29\\
        HP Cha               & 0.245 & 0.038 & 0.52 x 0.34 & 224 & 59.8 &  0.06 & 0.06 x 0.04 & 3.08\\
        HD 100453               & 0.915 & 0.103 & 0.37 x 0.27 & 281 & 209 &  0.36 & 0.06 x 0.05 & 2.65\\
        HD 135344B              & 0.324 & 0.044 & 1.2 x 0.31  & 336 & 574 &  0.16 & 0.09 x 0.08 & 3.06\\
        HD 143006               & 0.143 & 0.028 & 0.87 x 0.25 & 238 & 54.6 &  0.02 & 0.07 x 0.06 & 2.78\\
        RY Lup               & 0.348 & 0.042 & 0.46 x 0.27  & 350 & 280 &  0.14 & 0.08 x 0.06 & 3.59\\
        J1604         & 0.146 & 0.028 & 1.8 x 0.32  & 227 & 81.3 &  0.11 & 0.07 x 0.06 & 3.04\\
        J1608 & 0.198 & 0.029 & 0.85 x 0.34  & 335 & 119 &  0.42 & 0.1 x 0.09 & 2.79\\
        Sz111               & 0.224 & 0.031 & 0.72 x 0.36  & 335 & 160 &  0.41 & 0.1 x 0.08 & 2.52\\
        SR24S               & 0.811 & 0.102 & 0.533 x 0.154 & 108 & 30.8 &  0.06 & 0.07 x 0.04 & 2.60\\
        SR21               & 0.123 & 0.105 & 1.2 x 0.35 & 336 & 332 &  0.08 & 0.09 x 0.07 & 3.42\\
        DoAr44               & 0.333 & 0.047 & 1.2 x 0.35 & 336 & 172 &  0.08 & 0.09 x 0.08 & 2.10\\
        HD 169142               & 0.905 & 0.103 & 0.66 x 0.25  & 225 & 173 &  0.05 & 0.07 x 0.04 & 2.86\\ \bottomrule
\end{tabular}
    \end{adjustbox}
    {\centering \textbf{Notes:} \(^{a}\)Upper limits are 3\(\sigma\). \par}
\end{table*}

\subsection{ALMA}
The ALMA data for each source was selected from archival observations to both resolve the inner cavity of each transition disc and recover the largest spatial scales in the outer disc. The final calibrated visibilities for HD143006 were sourced from the DSHARP survey \citep{2018ApJ...869L..41A}. For all other discs, the highest resolution data was in ALMA bands 6 and 7 (1.3 and 0.88~mm, respectively), except for SR~24S, which has higher resolution observations in band 3 (3~mm). All ALMA data were reduced using the CASA pipeline for the appropriate ALMA cycle. Spectral lines in the ALMA data were flagged before extracting the continuum, which was time averaged to 30.5~s and a single channel per spectral window. ALMA visibilities were extracted using the \textit{exportuvfits} in CASA and subsequently dealt with in python, with the total integrated fluxes taken at zero spacing listed in Table~\ref{tab:3}.

\section{Modelling} \label{sect:model}
We quantify the relative position of the emission in the ATCA and ALMA observations using the code \textsc{Frankenstein} \citep{2020arXiv200507709J} to reconstruct the 1D radial brightness profiles of the disc at each wavelength. \textsc{Frankenstein} is an open source code that uses a Gaussian process to reconstruct the 1D radial brightness profile of a disc non-parametrically and can in principle measure radial features smaller than the clean beam size. We use an azimuthally symmetric brightness profile model despite known asymmetries in our sample as we are solely interested in obtaining the typical cavity radius for this exercise. This is especially true for ATCA observations with elongated beams, where the poor uv-coverage effectively results in different sensitivity and bias at baselines \(\rm \geq 300k\lambda\) and a 1D fit of the north-south/east-west elongation. While we would ideally prefer to construct 2D models, such modelling is not feasible due to the low resolution of our ATCA observations.

\textsc{Frankenstein} infers the brightness at a set of N radial points given a disc outer radius (\(\rm R_{max}\)) and assuming azimuthal symmetry. It then fits the profile to observed visibilities by using the discrete Hankel transform \citep{2015JOSAA..32..611B} to relate the observed visibilities to the radial brightness profile, applying a non-parametric Gaussian Process (GP) prior. The GP prior is learned from the data given two hyperparameters, \(\rm \alpha\) and \(\rm w_{smooth}\), and acts to dampen power in the reconstructed brightness profile on scales where the signal-to-noise in the visibilities is low. The most pertinent parameter is \(\rm \alpha\), which controls the signal-to-noise threshold below which \textsc{Frankenstein} does not attempt to fit the data. \(\rm w_{smooth}\) introduces a coupling between adjacent points and prevents regions of artificially low power arising from narrow gaps in the visibilities. For more details see \citet{2020arXiv200507709J}.
The brightness reconstruction is not overly sensitive to variations in \(\rm w_{smooth}\). For our sample we set \(\rm N=250\), \(\rm R_{max}\leqslant3^{\prime\prime}\) , \(\rm w_{smooth}\) to either 0.01 or 0.0001, and vary \(\rm \alpha\) between 1.01 and 1.5 and then select the model with the lowest \(\rm \chi^2\) value. However, if the data was sufficiently noisy at long baselines and the model was over-fitting, \(\rm \alpha\) was increased until oscillations in the fit at long baselines disappeared. We constrained our fits to be non-negative. We include Figure \ref{fig:param_sweep} as an example of the fitting methodology using a variety of hyperparamaters with the resulting \(\rm \chi^2\) values. Discs are deprojected using the inclination and position angle from high-resolution observations (see Table \ref{tab:4}). We restrict our modelling to sources with \(\sigma \geq 5\), limiting the number of sources in our sample that we model to CS Cha, HPCha, HD100453, HD135344B, RY Lup, J1608, Sz111, SR24S, DoAr44, and HD169142.

\section{Results} \label{sect:4}

\subsection{ATCA Continuum Emission}

We detected 14 out of the 15 discs from the survey with ATCA. HD34282 was the only non-detection and we present the $3\sigma$ flux upper limit in Table \ref{tab:3}. We spatially resolved the continuum emission of disc-like structures for 13 sources and show the CLEANed maps in Figure \ref{fig:all_images}. A deficit of emission is visible in the inner regions of HD100453, HD135344B, RY~Lup, J1608, Sz111, SR24S, DoAr44, and HD169142. This can be representative of a ring-like structure in the 8.8~mm grains of these discs assuming the bulk of the emission is from thermal dust. J1604 and SR21 also appear to exhibit a deficit of central emission, but due to the low SNR for both observations this is uncertain. The structure seen in the emission of HD135344B, RY Lup, J1604, J1608, and SR24S hints at asymmetrical dust structure which is expected to become more pronounced at longer wavelengths \citep{2013A&A...550L...8B, 2020arXiv201010568V}. However for RY~Lup this morphology is due to the disc being approximately edge-on \citep{2016ApJ...828...46A, 2018A&A...614A..88L}. The possible asymmetries in HD135344B and J1604 can not be confirmed due to the significant beam elongation, but it seems likely for HD135344B given the dust asymmetry in ALMA \citep{2016ApJ...832..178V}. The observations of HP~Cha, HD135344B, HD143006, J1604, and SR21 all suffered from a variety of poor weather conditions that required significant data flagging, resulting in sparse uv-coverage and low SNR. This makes it difficult to directly compare disc structures in the ATCA and ALMA continuum maps for these sources. 

\subsection{ALMA Degraded Continuum Emission}
In extended configurations of the 12-m array, ALMA has a resolutions range from \(\rm 0.020^{\prime\prime}\) at 230~GHz to \(\rm 0.043^{\prime\prime}\) at 110~GHz. To facilitate comparison between the ATCA and ALMA observations, we degrade the resolution of the ALMA observations to match that of the corresponding ATCA map using the \textit{restoringbeam} option in the \textit{tclean} task in CASA (c.f. column 4 of Table \ref{tab:3} and column 3 of Figure \ref{fig:all_images}).

The degraded ALMA continuum maps generally present morphology similar to the ATCA continuum maps. Two symmetric intensity peaks can be seen in the degraded ALMA images that align, within reasonable radial positional accuracy, for HD100453, RY~Lup, J1608, Sz111, SR21, DoAr44, and HD169142. The degraded ALMA maps for HD143006 and J1604 show morphology indicative of an inner cavity and ring-like structure. However, this is not reflected in the corresponding ATCA observations due to either the poor observing conditions or the lack of this structure in the 8.8~mm emission. For CS~Cha and HP~Cha, the degraded continuum maps exhibit elliptical structures similar to the corresponding ATCA observations. This suggests that any structure at 8.8~mm is not resolved by the clean reconstructed images. HD135344B exhibits two asymmetric intensity peaks indicative of its true asymmetric structure.

\begin{figure*}
  \centering
  \includegraphics[width=0.7\textwidth]{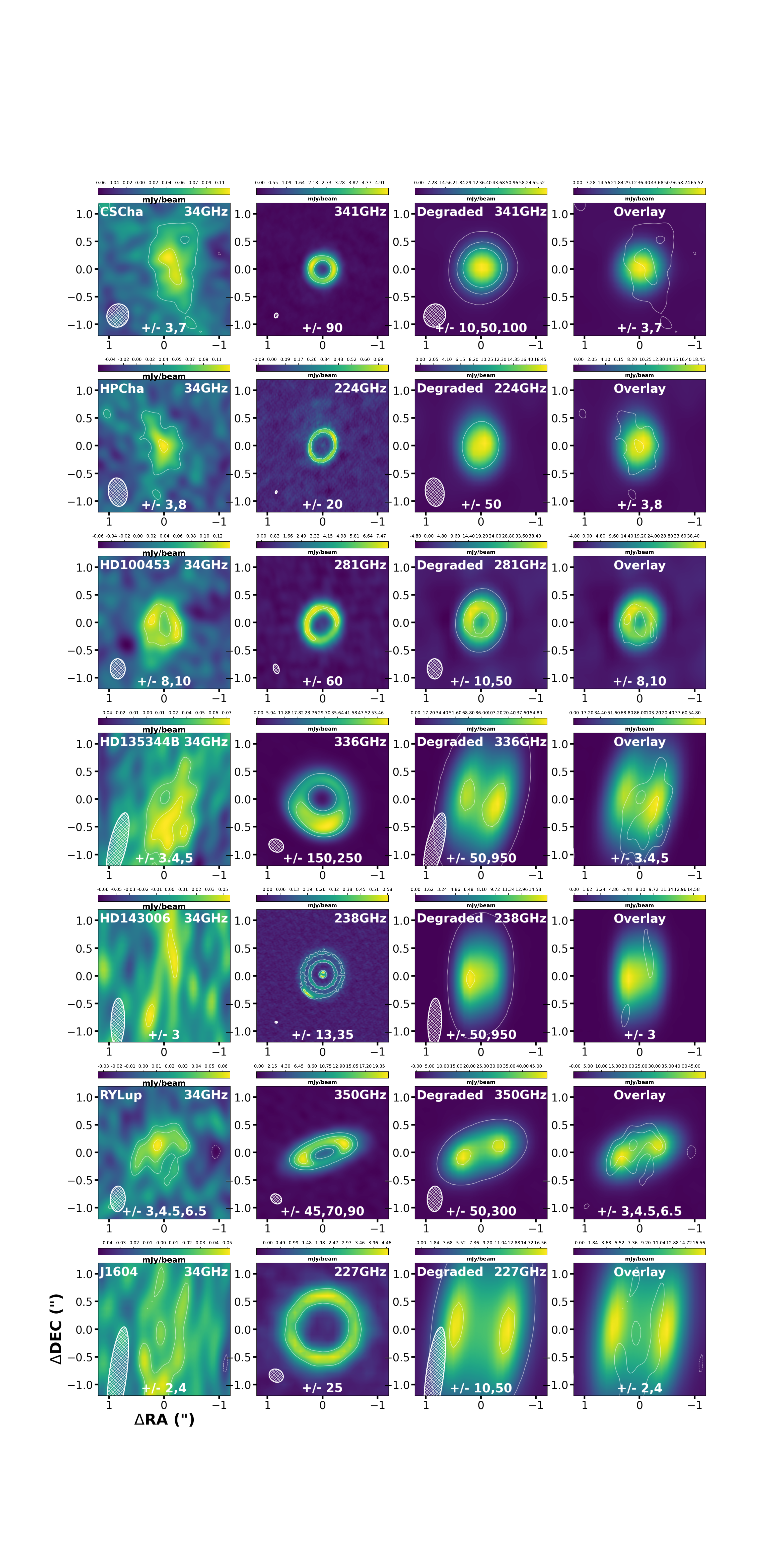}
  \caption{The 13 spatially resolved discs from the Discs Down Under survey. The $2^{\prime\prime}x2^{\prime\prime}$ maps show the 8.8~mm continuum ATCA maps, ALMA continuum maps, degraded ALMA continuum maps, and the ALMA continuum maps overlayed with ATCA continuum contours. The contours as printed at the bottom of each continuum map are a factor of the $\rm 1\sigma$ rms.}
\end{figure*}

\begin{figure*}
    \ContinuedFloat
  \includegraphics[width=0.7\textwidth]{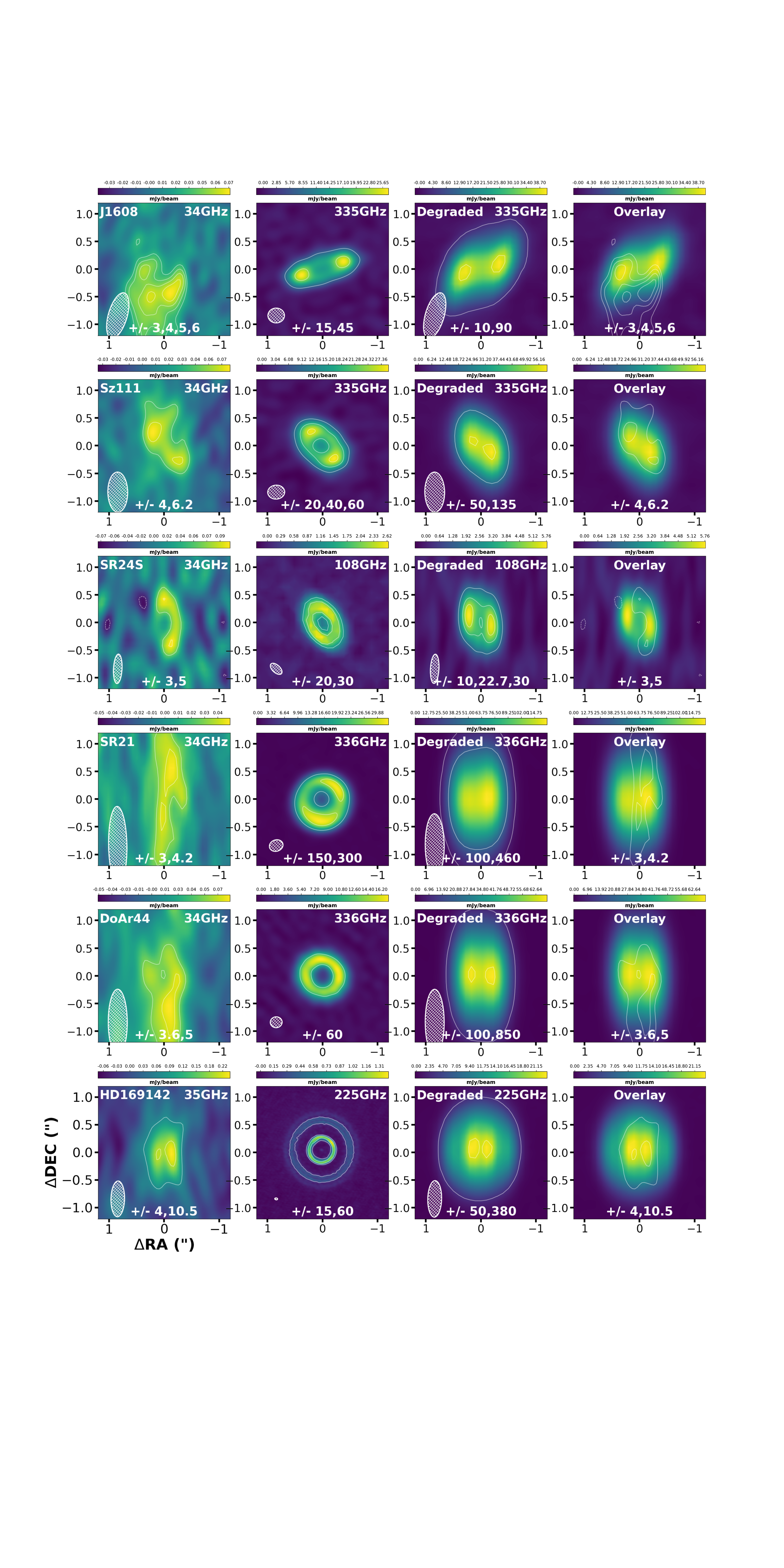}
  \caption{Continued.}
  \label{fig:all_images}
\end{figure*}

\subsection{Visibilities and Brightness Profiles}
\subsubsection{ATCA Phase Centering}
We estimate the phase centre of the ATCA observations where possible by centering the image to the apparent centre in the ring, which is taken as the central deficit of emission, for about half of the sample (HD100453, HD135344B, RY~Lup, J1608, Sz111, SR24S, DoAr44, and HD169142). For CS~Cha, the phase centre is taken as the stellar position correct by Gaia proper motions. HD135344B is known to be asymmetric \citep{2016ApJ...832..178V}, the position of central emission deficit closely corresponds to the orbital radius of the vortex derived by \citet{2018A&A...619A.161C} (81~au). For HP~Cha this comparison could not be made as it was too faint for Gaia to measure its proper motion. These phase centres are consistent (\(\rm \sim 0.2^{\prime\prime}\)) with the proper motion corrected Gaia positions and the accuracy is likely a function of the beam size, which can be worse in the north-south/east-west direction for sources with extended beams (e.g. the poor north-south accuracy for DoAr44). Offsets are similar to the astrometric accuracy seen in previous ATCA observations \citep{2015MNRAS.453..414W}.

The typical offsets of the phase centre ($\sim 0.2^{\prime\prime}$) are comparable to the cavity size of our transition discs. This directly impacts the \textsc{Frankenstein} 1D radial brightness models, in particular for sources where the 8.8~mm continuum map does not resolve an inner cavity with a ring of emission but exhibits asymmetrical structure. This means that the sources that appear centrally peaked rather than ring-like (CS~Cha and HP~Cha) might be asymmetric as well at 8.8~mm. Figure \ref{fig:ry_lup} highlights the effects on the 1D brightness models of RY~Lup when moving the phase centre, and cautions the interpretation of these models for sources that are possibly asymmetric (HP~Cha, HD135344B, RY~Lup, J1608, SR24S, and DoAr44) with relatively large beam sizes ($\gtrsim 0.2^{\prime\prime}$).

\subsubsection{Multi-Wavelength Comparison} \label{sect:multiwavelength_comparison}
\begin{figure*}
	\includegraphics[width=\textwidth]{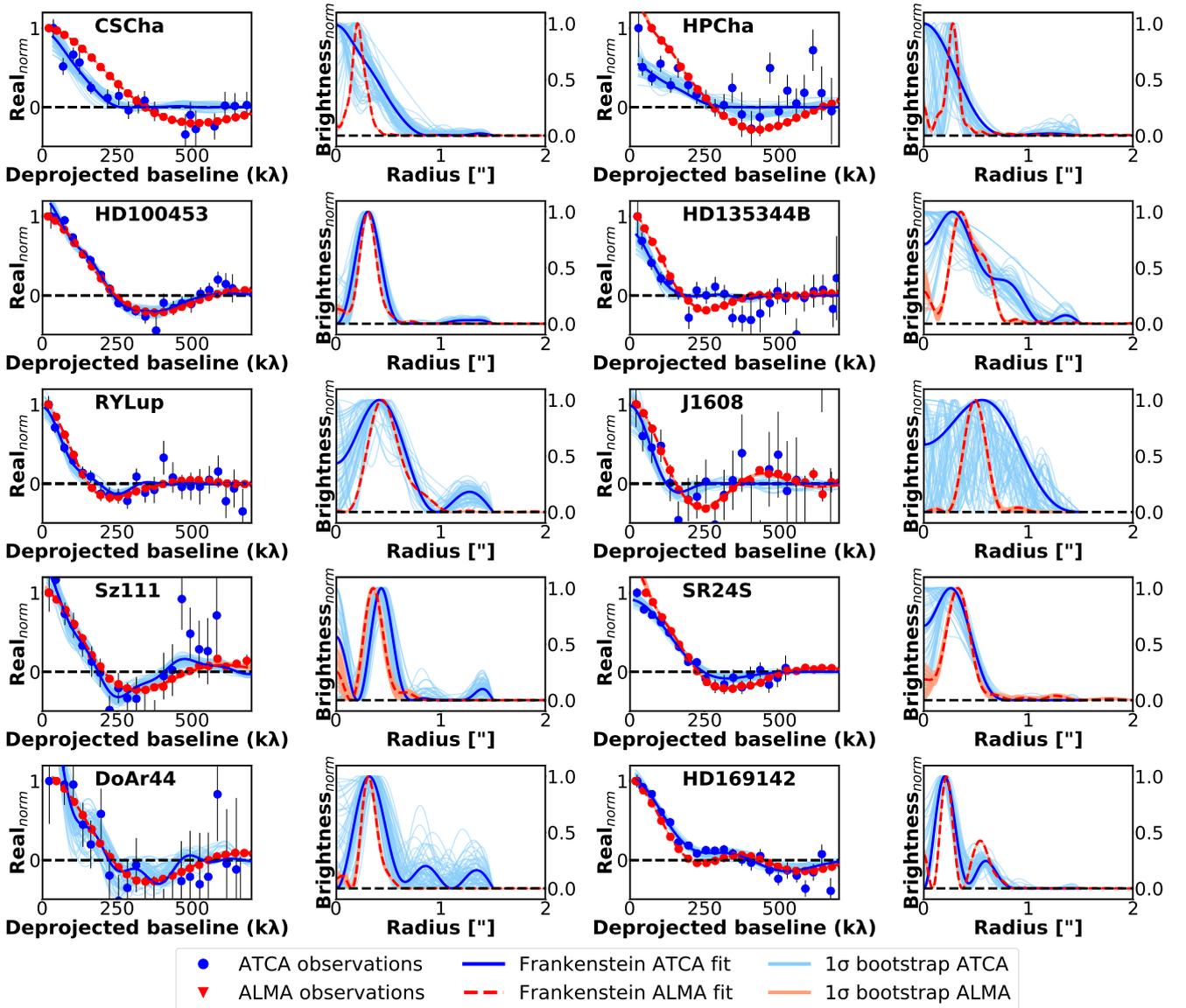}
	\caption{Normalised real component of the ATCA (blue) and ALMA (red) visibilities as a function of the deprojected baseline, and normalised \textsc{Frankenstein} model radial brightness profiles. The visibilities are overlaid by \textsc{Frankenstein} models (curves), and the model radial brightness profiles are shown with respective 1\(\sigma\) bootstrap error bar (see Sect. \ref{sect:multiwavelength_comparison} for further details on the error calculation)}.
	\label{fig:vis}
\end{figure*}

We present the real component of the deprojected ATCA and ALMA visibilities in Figure \ref{fig:vis}, adopting the inclination and position angle for each source from high-resolution observations (see Table \ref{tab:4}). The imaginary components of the visibilities are presented in Figure~\ref{fig:imag}. We also present in Figure~\ref{fig:vis} the \textsc{Frankenstein} reconstructed brightness profiles for the majority (10/15) of our sample and Table~\ref{tab:4} includes the radial offsets for the peak (col.2 and 3), \(\rm R_{peak}\) of the \textsc{Frankenstein} fits. Errors in the \textsc{Frankenstein} fit for sources with resolved features (e.g. a Gaussian peak) are calculated via the bootstrap method. This involves fitting the disc geometry as defined by the inclination and position angle, and calculating the model numerous times using different sub-samples of the data. The bootstrap method effectively tests the 'goodness' of a fits chosen geometry and hyperparameters. As a result, if the data has a relatively high SNR and the choice of hyper-parameters is correct, sub-samples of the data will always reproduce approximately the same fit. This is apparent in our ALMA visibility fits where the \(\rm 1\sigma\) error is \(\rm \sim 0.4\% \) of the \(\rm R_{peak}\) values. For further details on the bootstrap method in \textsc{Frankenstein} see \citet{2020arXiv200507709J}. We take the uncertainty in the \(\rm R_{peak}\) values as the combination of the \(\rm 1\sigma\) standard deviation of our bootstrap models and the full width half maximum of the observational beam size. For sources with unresolved features, calculating \(\rm R_{peak}\) and the fwhm is not possible, and the results for these sources are omitted from Table \ref{tab:4}. We include radial offsets from previous visibility fits in the literature and note that our values are comparable (see Table~\ref{tab:4}, col.4). The visibilities for ATCA observations that were not modelled with \textsc{Frankenstein} (HD34282, SZ~Cha, HD143006, J1604, and SR21) and the corresponding ALMA visibilities with \textsc{Frankenstein} fits are included in Appendix \ref{app:other_vis}.

Our multi-wavelength visibility comparison shown in Figure \ref{fig:vis} suggests that the 8.8~mm and sub-mm disc structures share a similar cavity size for HD100453, HD135344B, RY~Lup, J1608, Sz111, SR24S, DoAr44, HD169142. This is indicated by the similar location of the null seen in the visibility data and similar radial positions of the Gaussian ring peak seen in the \textsc{Frankenstein} reconstructed 1D brightness profiles. For CS~Cha and HP~Cha the 1D radial brightness profiles suggest that the 8.8~mm and sub-mm do not share a similar cavity size. For these sources, it is clear that the noise is larger than the correlated flux of the negative part in the ALMA data. It is unlikely that the \textsc{Frankenstein} fit for low SNR at long baselines could resolve a cavity if one is indeed present. The ATCA 1D radial brightness profiles of HD100453, HD135344B, RY~Lup, J1608, Sz111, SR24S, and DoAr44 show an inner deficit of emission consistent with Figure \ref{fig:all_images}. For the ATCA data of HD169142 the \textsc{Frankenstein} fit is able to reproduce the two rings seen in the ALMA image that is not resolved in the ATCA image in Figure \ref{fig:all_images}, this is attributed to \textsc{Frankenstein's} ability to resolve angular scales smaller than the uniform-weighted clean beam for sources with relatively low SNR at long baselines. All of the ALMA radial brightness profiles indicate a ring-like structure, as also shown in Figure \ref{fig:all_images}.

\begin{table*}
 \centering
\caption{Radial brightness profile properties of the ATCA and ALMA images for sources modelled with \textsc{Frankenstein}. This includes the radial offset of the peak brightness \(\rm R_{peak}\) (col. 2 and 3). We compare our \textsc{Frankenstein} fits with previous visibility models in col.4. Also included are the 1D profile peak positions from other models in the literature from infrared (IR) and \(\rm 13CO\) observations, and the observed inclination and position angle of the target disc. Superscripts in the column headers refer to the position in the reference list of the last column.}
\label{tab:4}
\begin{adjustbox}{width=1\textwidth}
\begin{tabular}{@{}ccccccccc@{}}
\toprule
Source & \begin{tabular}[c]{@{}c@{}}$R_{\rm peak}$\\ ATCA obs\\ (")\end{tabular} & \begin{tabular}[c]{@{}c@{}}$R_{\rm peak}$\\ ALMA obs\\ (")\end{tabular} & \begin{tabular}[c]{@{}c@{}}$R_{\rm peak}$\\ \(\rm ALMA^{a}\) \\ (")\end{tabular} & \begin{tabular}[c]{@{}c@{}}$R_{\rm peak}$\\ \(\rm IR^{b}\) \\ (")\end{tabular} & \begin{tabular}[c]{@{}c@{}}$R_{\rm peak}$\\ \(\rm 13CO^{c}\) \\ (")\end{tabular} & \begin{tabular}[c]{@{}c@{}}\(\rm Incl.^{d}\) (\(\degree\)) \end{tabular}  & \begin{tabular}[c]{@{}c@{}}\(\rm PA^{d}\) (\(\degree\))  \end{tabular} & \begin{tabular}[c]{@{}c@{}}Ref. \\ (a,b,c,d) \end{tabular} \\ \midrule
CS Cha & <0.390\(^{u}\) & \(0.204_{-0.007}^{+0.007}\) & - & 0.102 & - &  8 & 161 & -, 8, -, 5 \\ \\
HP Cha & <0.340\(^{u}\) & \(0.277_{-0.007}^{+0.007}\) & - & - & - &  37 & 164 & -, -, -, 6 \\ \\
HD100453 & \(0.307_{-0.115}^{+0.115}\) & \(0.306_{-0.009}^{+0.009}\) & \(0.24_{-0.01}^{+0.01}\) & 0.184 & - &  29.5 & 151 & 3, 9, -, 3 \\ \\
HD135344B & \(0.278_{-0.155}^{+0.155}\) & \(0.356_{-0.014}^{+0.014}\) &  \(0.402_{-0.001}^{+0.001}\) & 0.363 & 0.292 &  17.7 & 62 & 1, 10, 16, 6 \\ \\
RY Lup & \(0.412_{-0.190}^{+0.190}\) & \(0.436_{-0.011}^{+0.011}\) & \(0.452_{-0.009}^{+0.009}\) & - & 0.286 &  68 & 109 & 1, -, 14, 7 \\ \\
J1608 & \(0.557_{-0.269}^{+0.269}\) & \(0.492_{-0.016}^{+0.016}\) & \(0.389_{-0.012}^{+0.018}\) & 0.288 & 0.300 &  -74 & 107 & 1, 11, 14, 8 \\ \\
Sz111 & \(0.432_{-0.189}^{+0.189}\) & \(0.356_{-0.015}^{+0.015}\) & \(0.282_{-0.015}^{+0.020}\) & - & 0.225 & -53 & 40 & 1, -, 14, 7 \\ \\
SR24S & \(0.263_{-0.123}^{+0.123}\) & \(0.330_{-0.013}^{+0.013}\) & \(0.304_{-0.001}^{+0.001}\) & - & - & 46 & 25 & 4, -, -, 1 \\ \\
DoAr44 & \(0.317_{-0.163}^{+0.163}\) & \(0.310_{-0.014}^{+0.014}\) & \(0.285_{-0.001}^{+0.001}\)  & 0.123 & 0.146 & 20 & 30 & 1, 12, 16, 1 \\ \\
HD169142 & \(0.203_{-0.089}^{+0.089}\), & \(0.216_{-0.010}^{+0.010}\), & \(0.235_{-0.034}^{+0.034}\), & 0.118 & 0.121  & 13 & 5 & 2, 13, 15, 2 \\ \\
& \(0.591_{-0.099}^{+0.099}\)* & \(0.583_{-0.017}^{+0.017}\)* & \(0.594_{-0.034}^{+0.034}\)*  & - & -  \\\bottomrule
\end{tabular}
\end{adjustbox}
    {\centering \textbf{Notes:} \(^{u}\) Upper limits are taken as the maximum cavity size (if a cavity exists) from the error for each source in Figure \ref{fig:r_peak_plot}. \\ *Second peak. \\ \textbf{References} (1) \citet{2018ApJ...859...32P}, (2) \citet{2017A&A...600A..72F}, (3) \citet{2019A&A...624A..33V}, (4) \citet{2017ApJ...839...99P}, (5) \citet{2020francis} (6) \citet{2018A&A...619A.161C}, (7) \citet{2018ApJ...859...21A}, (8) \citet{2018A&A...616A..79G}, (9) \citet{2017A&A...597A..42B}, (10) \citet{2017ApJ...849..143S}, (11) \citet{2019A&A...624A...7V}, (12) \citet{2018MNRAS.477.5104C}, (13) \citet{2019MNRAS.486.3721B}, (14) \citet{2018ApJ...854..177V}, (15) \citet{2018A&A...614A.106C}, (16) \citet{2016A&A...585A..58V}. \par}
\end{table*}

\subsection{The Spectral Slope}
\begin{figure*}
	\includegraphics[width=0.99\textwidth]{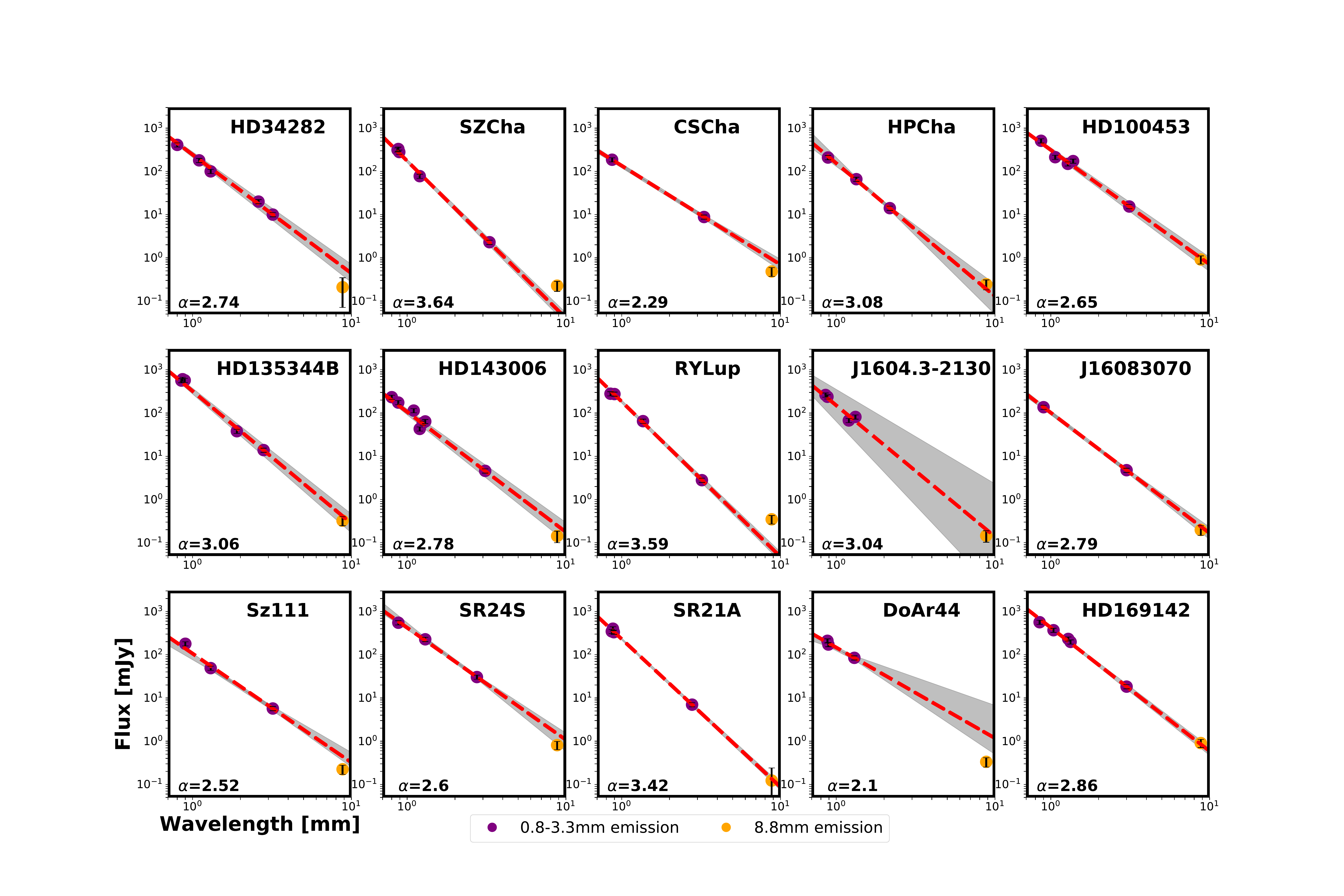}
	\caption{Spectral slopes for 14 sources in the Discs Down Under survey with observed emission between 0.8 and 3.3~mm. The error bars shown in both panels includes a calibration uncertainty of 10\% (ALMA) and 20\% (ATCA). Over-plotted with grey shading are linear least-squares fits error limits to parts of the spectral slope between 0.8 and 3.3~mm.}
	\label{fig:seds}
\end{figure*}

In our multi-wavelength comparison we have assumed that the 8.8~mm emission is primarily from thermal dust, which should be tested. Figure \ref{fig:seds} shows the spectral slope between 0.8~mm and 1~cm for sources with sufficient flux measurements between 0.8--3.3~mm for a linear least-squares fit (in log-log space). We include a calibration uncertainty of 10\% for ALMA data and 20\% for ATCA data. Figure \ref{fig:seds} shows that for the majority of our sources the 8.8~mm emission falls within the fitted error, indicating that the emission is likely due to thermal dust emission from large grains. For SZ~Cha and RY~Lup, we see excess emission above that of thermal dust, which is likely due to free-free emission. The lack of publicly available ALMA Band 3 data for J1604 and DoAr44 result in large error bands in the spectral slope. In Table \ref{tab:3} we include the spectral slope (\(\rm \alpha_{mm}\)) calculated from our linear least-squares fit. The spectral indices for our sample range from \(\rm 2.09 - 3.63\). This is consistent with a distribution of large grains \citep{2004ASPC..323..279N, 2006ApJ...636.1114D}. 

\section{Discussion} \label{sect:5}
\subsection{Other Sources of Emission} \label{sect:other_emission}
Continuum emission at millimetre wavelengths from discs around young stars often show excess emission above that expected from a simple extrapolation of thermal dust emission observed at shorter mm wavelengths \citep{2006A&A...446..211R, 2009A&A...495..869L, 2012MNRAS.425.3137U, 2017MNRAS.466.4083U}. This excess has been attributed to thermal free-free emission from an ionised wind, non-thermal processes such as chromospheric emission from the young stellar object, or a combination of both \citep{2007prpl.conf..555D,2007prpl.conf..539M}. For thermal free-free emission, the total integrated flux can vary by a factor of \(\rm 20-40\)\% over a long period, while for non-thermal emission the total integrated flux can vary by up to a factor \(\gtrsim 2\) on very shorter times scales (minutes to hours) \citep{2012MNRAS.425.3137U}. In our sample, only SZ~Cha and RY~Lup show an excess in emission above that expected from thermal dust. This excess is less than 40\% of the total integrated flux and is likely due to free-free emission. However, further observations at mm-wavelengths over a range of candences are required to confirm this. Previous temporal flux monitoring by \citet{2012MNRAS.425.3137U} also found that RY~Lup exhibits excess emission. Free-free emission has been observed in a number of other transition discs not in our sample, including HD97048 \citep{2017A&A...597A..32V}, HD142527 \citep{2015ApJ...812..126C}, and HD100546 \citep{2015MNRAS.453..414W}. We expect the emission from the bulk of our sample to indeed be due to thermal dust, and hence the emission comparison with ALMA emission is likely a dust-to-dust comparison.

\subsection{Dust Traps and Cavity Formation}
\begin{figure}
	\includegraphics[width=0.48\textwidth]{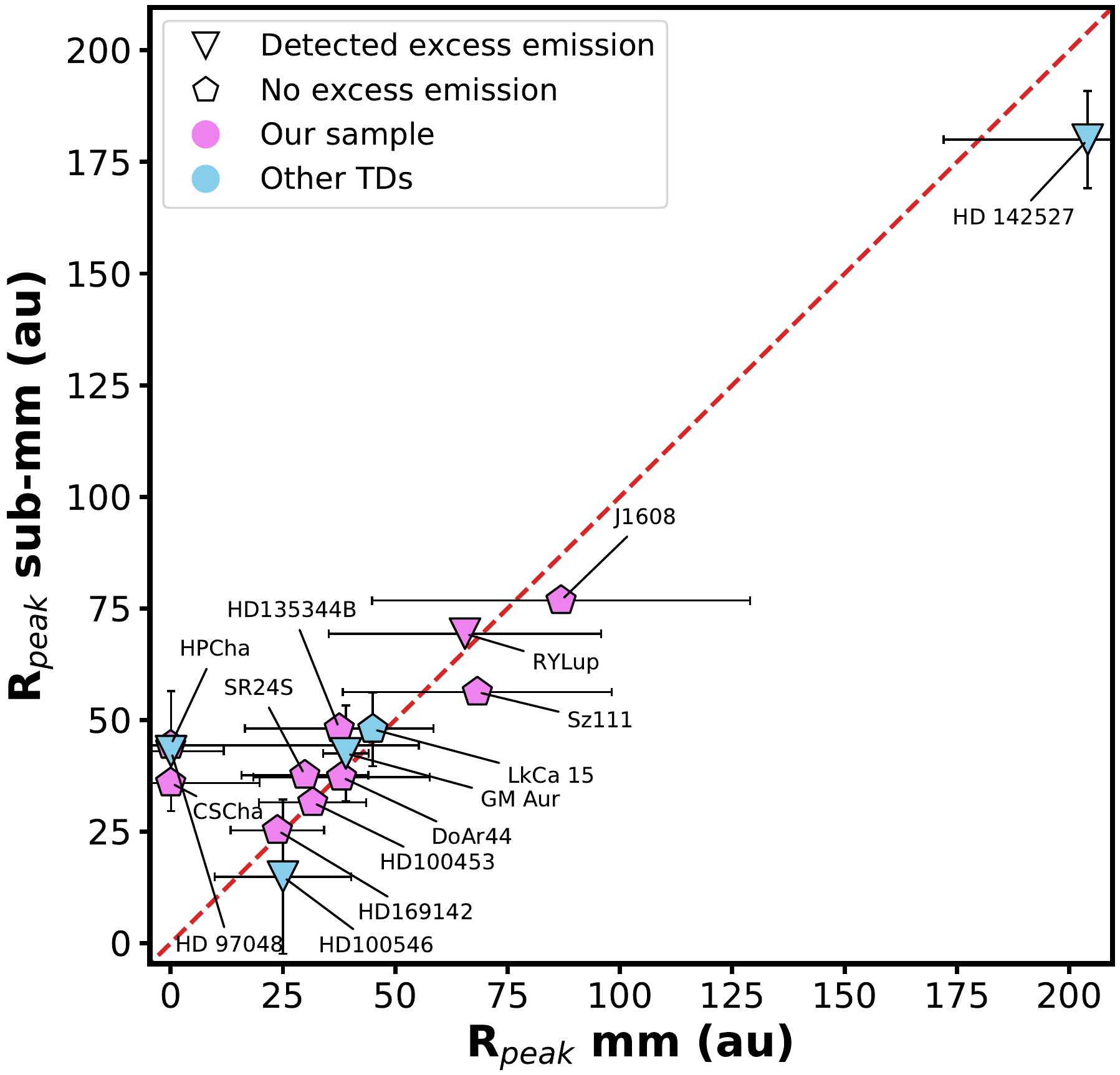}
	\caption{The $\rm0.43-1.3mm$ $\rm R_{peak}$ radial offsets as a function of the $\rm7-9mm$ $\rm R_{peak}$ radial offsets. The dashed line represents an equal radial offset at both wavelength ranges. The pentagon and triangle markers indicate whether emission other than thermal dust emission has been detected in the source as no excess or detected excess respectively. The pink markers represent $\rm R_{peak}$ values from our \textsc{Frankenstein} modelling, and the blue markers represent $\rm R_{peak}$ values from the literature including: HD100546 $\rm0.86~mm$ data from \citet{2018ApJ...859...32P} and $\rm7~mm$ data from \citet{2015MNRAS.453..414W}; GM Aur $\rm0.9~mm$ and $\rm7~mm$ data from \citet{2018ApJ...865...37M}; LkCa 15 $\rm0.43~mm$ data from \citet{2018ApJ...859...32P} and $\rm7~mm$ data from \citet{2014ApJ...788..129I}; HD142527 $\rm1.3~mm$ data from \citet{2020arXiv201015310G} and $\rm7~mm$ data from \citet{2015ApJ...812..126C}; and HD97048 $\rm0.85~mm$ and $\rm8.9~mm$ data from \citet{2017A&A...597A..32V}. The errors are a propagation of the bootstrap \textsc{Frankenstein} errors and the full width half maximum of the beam size.}
	\label{fig:r_peak_plot}
\end{figure}

The 1D radial brightness profile models derived by \textsc{Frankenstein} highlight that the majority of our sample (HD100453, HD135344B, RY~Lup, J1608, Sz111, SR24S, DoAr44, and HD169142) share a similar $\rm R_{peak}$ in both the ATCA and ALMA data. The models exhibit a Gaussian ring fit indicative of a resolved inner cavity in Fourier space that is likely due to the concentration of grains in dust traps and hence, we take $R_{\rm peak}$ as representative cavity size. Neither CS~Cha or HP~Cha share similar $\rm R_{peak}$ values between the ATCA and ALMA data. For these sources, the ATCA 1D radial brightness profiles suggest a disc without an inner deficit of emission, and the ALMA fits suggest a ring-like morphology. As a result, if we plot $\rm R_{peak}$(ALMA) versus $\rm R_{peak}$(ATCA) (see Figure \ref{fig:r_peak_plot}), our sample of transition discs can be classified into two groups: sources with similar $\rm R_{peak}$ values and those which only show a cavity in the ALMA data. Sources in the first group from our sample include HD100453, HD135344B, RY~Lup, J1608, Sz111, SR24S, DoAr44, and HD169142. Other transitions discs in the literature which exhibit an approximately one-to-one correlation between the 8.8~mm and sub-mm $\rm R_{peak}$ values include HD100546 \citep{2015MNRAS.453..414W, 2018ApJ...859...32P}, GM~Aur \citep{2018ApJ...865...37M}, LkCa15 \citep{2014ApJ...788..129I, 2018ApJ...859...32P}, and HD142527 \citep{2015ApJ...812..126C}. The second group includes CS~Cha and HP~Cha from our sample, and HD97048 from \citet{2017A&A...597A..32V}. For the purposes of this paper, group 1 sources will be labelled resolved cavity transition discs (RC-TDs) and group 2 sources will be labelled non-detected cavity transition discs (NC-TDs).

The non-detection of cavities at 8.8\,mm for the NC-TDs can reflect the lack of a cavity in the large grain population, but could also be due to large asymmetric structure resulting in either a poor model fit or inaccurate phase centering, unresolved features due to spatial resolution limitations, or the presence of other sources of emission obscuring disc features as suggested by \citet{2017A&A...597A..32V} for HD97048. The emission for CS~Cha and HP~Cha appears to be dominated by thermal dust (see Figure \ref{fig:seds}). Due to the lack of proper motions and the astrometric uncertainty of ATCA, it is possible that the emission of HP~Cha is asymmetric and the chosen phase centre is incorrect (e.g. see Figure \ref{fig:ry_lup}), as asymmetries are expected to become more pronounced or even only revealed at longer wavelengths \citep{2013A&A...550L...8B, 2020arXiv201010568V}. However, it is equally likely, given the low spatial resolution of the ATCA observations, that a cavity exists for CS~Cha and HP~Cha but remains unresolved. Using \textsc{Frankenstein} to fit low SNR visibilities will result in poor fits not representative of the true disc features. As a result, the presence of a cavity at 8.8~mm cannot be confirmed for CS~Cha and HP~Cha, and more observations are required.

The exact mechanism driving large cavity formation in transition discs remains unknown \citep{2011ARA&A..49...67W}. However historically the observed deficit of inner disc emission has been attributed to either grain growth, photoevaporation, or the trapping of dust due to either dead zones or the dynamical interaction with a companion. Grain growth models \citep{2008A&A...487..265L, 2012A&A...544A..79B, 2015P&SS..116...48G} predict, due to the settling, growth, and inward drift of grains, that the disc will contain a size-sorted radial distribution of dust, with larger grains preferentially concentrated towards the mid-plane and host star. Previous reviews of photoevaporative models \citep{2017RSOS....470114E} conclude that the deficit of inner emission seen in transition discs with large cavities (\(\rm \geq 20au\)) and mass accretion rates \(\rm \geq 10^{-9} M_{\odot} yr^{-1}\) cannot be attributed entirely to photoevaporation. Therefore, we suggest that neither grain growth nor photoevaporation are likely to be dominating the grain evolution in our RC-TD sources. Our sample does not show evidence of radial size sorting (Figure 2), exhibit large cavities (Figure \ref{fig:r_peak_plot}), and the majority have modest levels of accretion (Table \ref{tab:1}).

Magnetohydrodynamical (MHD) models of transition discs from \citet{2016A&A...596A..81P} predict a factor of 2 difference in the radial peak positions for sub-mm to mm grains at dust traps located at the edge of dead zones \citep[see Figure 4 at 1Myr in][]{2016A&A...596A..81P} which is excluded by our one-to-one correlation seen in Figure \ref{fig:r_peak_plot}. Additionally, MHD models require a high turbulence viscosity coefficient (\(\rm \alpha \approx 10^{-2}\)) in active regions of the disc \citep{2016A&A...596A..81P, 2019ApJ...871...10U}, which is inconsistent with observations of classical protoplanetary discs \citep[e.g. \(\rm \alpha \approx 10^{-3} - 10^{-4}\) from][]{2016ApJ...816...25P,2020ApJ...895..109F} and previous studies of transition discs that model how inner companions influence cavity formation \citep{2017ApJ...843..127D, 2020ApJ...888L...4T}. \citet{2017ApJ...843..127D} showed with hydrodynamical simulations and radiative transfer models that the turbulence viscosity coefficient for HD169142 had to be lower than \(10^{-4}\) for a single planet to form two distinct ring that agreed with the observations. However \citet{2020ApJ...888L...4T}, using a similar procedure but with two planets, were able to reproduce the two observed rings with $\rm \alpha=5\times10^{-3}$. Turbulence in the active regions of transition discs may not be well enough constrained to reach a firm conclusion regarding its role in cavity formation.

The dynamical interaction with a companion is proposed to be the driving force behind most features seen in transition discs including the formations of dust traps. For our RC-TDs, \citet{2015A&A...579A.106V, 2016A&A...585A..58V, 2018ApJ...854..177V} showed gas cavities inside dust cavities for SR21, HD135344B, DoAr44, J1604, Sz111, J16083070. \citet{2019A&A...624A..33V} utilised SPH and radiative transfer codes to show some disc features seen in HD100453 can be attributed to an undetected, low mass close companion within the disc's cavity. The study of HD100453 is further refined by \citet{Gonzalez2020} and \citet{Nealon2020}, who show that while the outer disc morphology can be caused by a companion star on an inclined orbit exterior to the disc, the inner cavity can be explained by a $\rm \lesssim5$~M$_\mathrm{J}$ planet, less massive than previously suggested \citep{2019A&A...624A..33V}, at 15-20~au. Using similar 1D radial brightness profiles for the majority of our sources (HD135344B, RY Lup, J1608, Sz111, SR24S, and DoAr44), \citet{2017ApJ...839...99P} and \citet{2018ApJ...859...32P} suggest that the clearing of the inner cavity could be attributed to the dynamical interaction with an embedded planet. For HD169142, \citet{2017A&A...600A..72F} uses a thermo-chemical code to model the gas and dust distribution. Their results suggest that dynamical interaction between the disc and two giant embedded planets results in the depletion of the inner disc, and creates two pressure bumps that aid in the trapping of dust at each observable ring position. It still remains unclear for the majority of transition discs if a companion is indeed responsible for the formation of the cavity, whether this companion is an embedded planet or a binary star. What is notable about our sample is that for the RC-TDs the 8.8~mm and sub-mm grains share a similar cavity size ($\rm R_{peak}$ value). This is also seen in SR24S and HD142527 (included in Figure \ref{fig:r_peak_plot}). \citet{2019ApJ...878...16P} presents a planet-disc model for SR24S that predicts both mm and sub-mm grains will share a similar radial peak in the dust density. \citet{2018MNRAS.477.1270P} extensively modelled HD142527 with 3D hydrodynamical simulations considering several possible orbits for the M-dwarf binary companion presented in \citet{2016A&A...590A..90L}. They conclude that all the observable disc features can be attributed to the tidal truncation of the disc from the binary companion. Given that our RC-TDs and HD142527 follow the one-to-one correlation shown in Figure \ref{fig:r_peak_plot}, we suggest similar physical mechanisms affecting the grain evolution for these discs may have occurred. 

We include in Table \ref{tab:4} the IR and \(\rm ^{13}CO\) $\rm R_{peak}$ values where possible for our sample. The IR and \(\rm ^{13}CO\) $\rm R_{peak}$ values are consistently smaller than the corresponding ATCA/ALMA $\rm R_{peak}$ values, and for sources with both IR and \(\rm ^{13}CO\) observations, the respective $\rm R_{peak}$ values are similar. The discrepancy between the ATCA and ALMA versus IR and\(\rm ^{13}CO\) $\rm R_{peak}$ values was originally classified as the "missing cavities" problem by \citet{2012ApJ...750..161D}. With ATCA and ALMA observations tracing large dust grains (sub-mm to mm sizes) and IR observations tracing small dust grains (\(\rm \sim \mu m\) sizes) our comparison is in agreement with dust evolution models which show large dust grains  accumulate at pressure maxima (dust traps), while smaller dust grains are coupled to the gas and follow the accretion flow into the inner disc \citep{2012A&A...539A.148B, 2012A&A...545A..81P, 2012ApJ...755....6Z}.

\section{Conclusions} \label{sect:6}
In this work we  present the largest 8.8~mm continuum survey to date of transition discs with large cavities, and compare the resulting dust emission to ALMA observations.

\begin{enumerate}
\item For the 15 discs observed with ATCA, 14 were detected, 13 were spatially resolved, and 8 exhibited morphology indicative of ring-like structure.

\item The spectral slopes indicate that the emission is dominated by thermal dust for most sources, suggesting the ATCA emission is tracing the large grain population. Only SZ Cha and RY Lup showed excess emission above that from thermal dust emission. 

\item We use the Frankenstein code to model 1D radial brightness profiles, which reveal that both the 8.8~mm ATCA and sub-mm ALMA emission in most of the discs share a similar  radial position of the peak brightness, \(\rm R_{peak}\), suggesting a similar cavity size.

\item The \(\rm R_{peak}\) values for the millimetre and sub-mm data is consistently larger than the corresponding IR and \(\rm ^{13}CO\) \(\rm R_{peak}\) values. This is in agreement with models indicating large grains preferentially concentrating at maxima in the gas pressure, whilst small \(\rm \sim \mu m\)-sized grains are coupled to the gas.

\item We suggest that the large cavities in these transition discs result from a dust trap, likely induced by a companion.

\end{enumerate}

\section*{Acknowledgements}
The authors thank the referee for their constructive comments and suggestions. We thank Mathew Agnew and Elodie Thilliez for useful discussions with our ATCA observing proposal. B.J.N is supported by an Australian Government Research Training Program (RTP) Scholarship.  CP acknowledges funding from the Australian Research Council via grants FT170100040 and DP180104235. N.M. acknowledges support from the Banting Postdoctoral Fellowships program, administered by the Government of Canada. RAB acknowledges support from the STFC consolidated grant ST/S000623/1. This work has also been supported by the European Union's Horizon 2020 research and innovation programme under the Marie Sklodowska-Curie grant agreement No 823823 (DUSTBUSTERS). J.-F.G. acknowledges funding from ANR (Agence Nationale de la Recherche) of France under contract number ANR-16-CE31-0013 (Planet-Forming-) and thank the LABEX Lyon Institute of Origins (ANR-10-LABX-0066) of the Universit\'e de Lyon for its financial support within the programme `Investissements d'Avenir' (ANR-11-IDEX-0007) of the French government operated by the ANR. C. M. Wright acknowledges financial support from the Australian Research Council via Future Fellowship FT100100495. The Australia Telescope Compact Array is part of the Australia Telescope which is funded by the Commonwealth of Australia for operation as a National Facility managed by CSIRO. This research has made use of NASA's Astrophysics Data System. The National Radio Astronomy Observatory is a facility of the National Science Foundation operated under agreement by the Associated Universities, Inc. ALMA is a partnership of ESO (representing its member states), NSF (USA) and NINS (Japan), together with NRC (Canada) and NSC and ASIAA (Taiwan) and KASI (Republic of Korea), in cooperation with the Republic of Chile. The Joint ALMA Observatory is operated by ESO, AUI/ NRAO and NAOJ. This paper makes use of the following ALMA data: ADS/JAO.ALMA\#2012.1.00158.S, ADS/JAO.ALMA\#2012.1.00761.S, ADS/JAO.ALMA\#2013.1.00437.S, ADS/JAO.ALMA\#2015.1.00192.S, ADS/JAO.ALMA\#2015.1.00888.S, ADS/JAO.ALMA\#2015.1.01353.S, ADS/JAO.ALMA\#2016.1.00344.S, ADS/JAO.ALMA\#2016.1.00484.L, ADS/JAO.ALMA\#2017.1.00449.S, ADS/JAO.ALMA\#2017.1.00884.S, ADS/JAO.ALMA\#2017.1.00969.S, ADS/JAO.ALMA\#2017.1.01424.S, ADS/JAO.ALMA\#2017.1.01460.S

This work makes use of the following software: The Miriad package \citep{1995ASPC...77..433S}, the Common Astronomy Software Applications (CASA) package \citep{2007ASPC..376..127M}, Python version 3.7, astropy \citep{2013A&A...558A..33A}, matplotlib \citep{Hunter:2007}, and \textsc{Frankenstein} \citep{2020arXiv200507709J}.
 
\section*{Data Availability}
The ATCA observational data used in this paper is available from the Australian National Telescope Facility Archive at \url{https://atoa.atnf.csiro.au/} under project code C3119. The ALMA observational data is available from the ALMA science archive at \url{https://almascience.nrao.edu/aq/} under the project IDs listed in Table \ref{tab:1}. 1D radial brightness models used the Frankenstein code which is available from \url{https://github.com/discsim/frank}. Reduced observation data and models will be shared on reasonable request to the corresponding author.




\bibliographystyle{mnras}
\bibliography{mnras} 



\appendix

\section{HD163296} \label{app:hd163296}
\begin{figure}
	\includegraphics[width=0.48\textwidth]{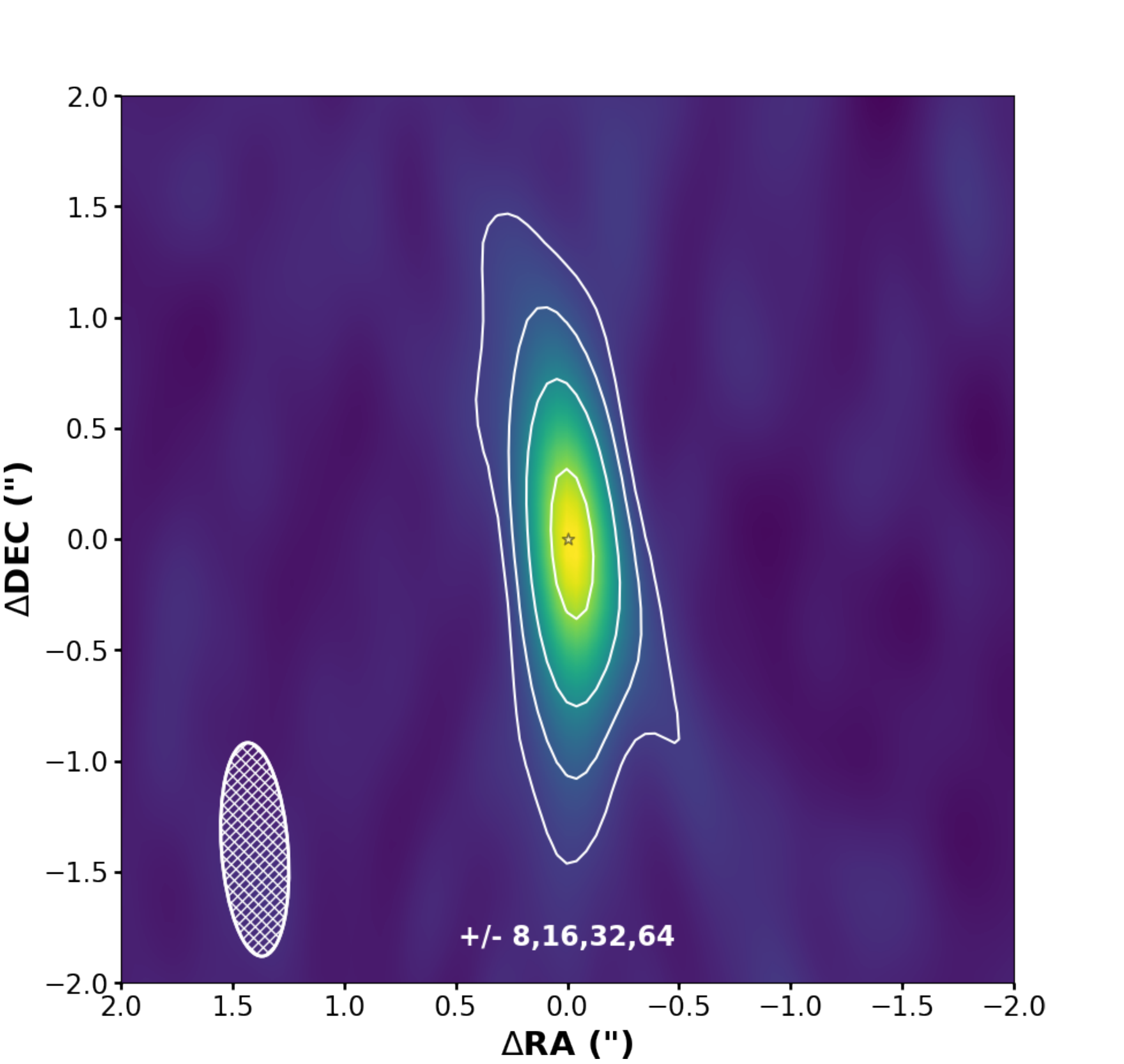}
	\caption{8.8~mm continuum map of HD163296.}
	\label{fig:hd163296}
\end{figure}

\begin{figure}
	\includegraphics[width=0.48\textwidth]{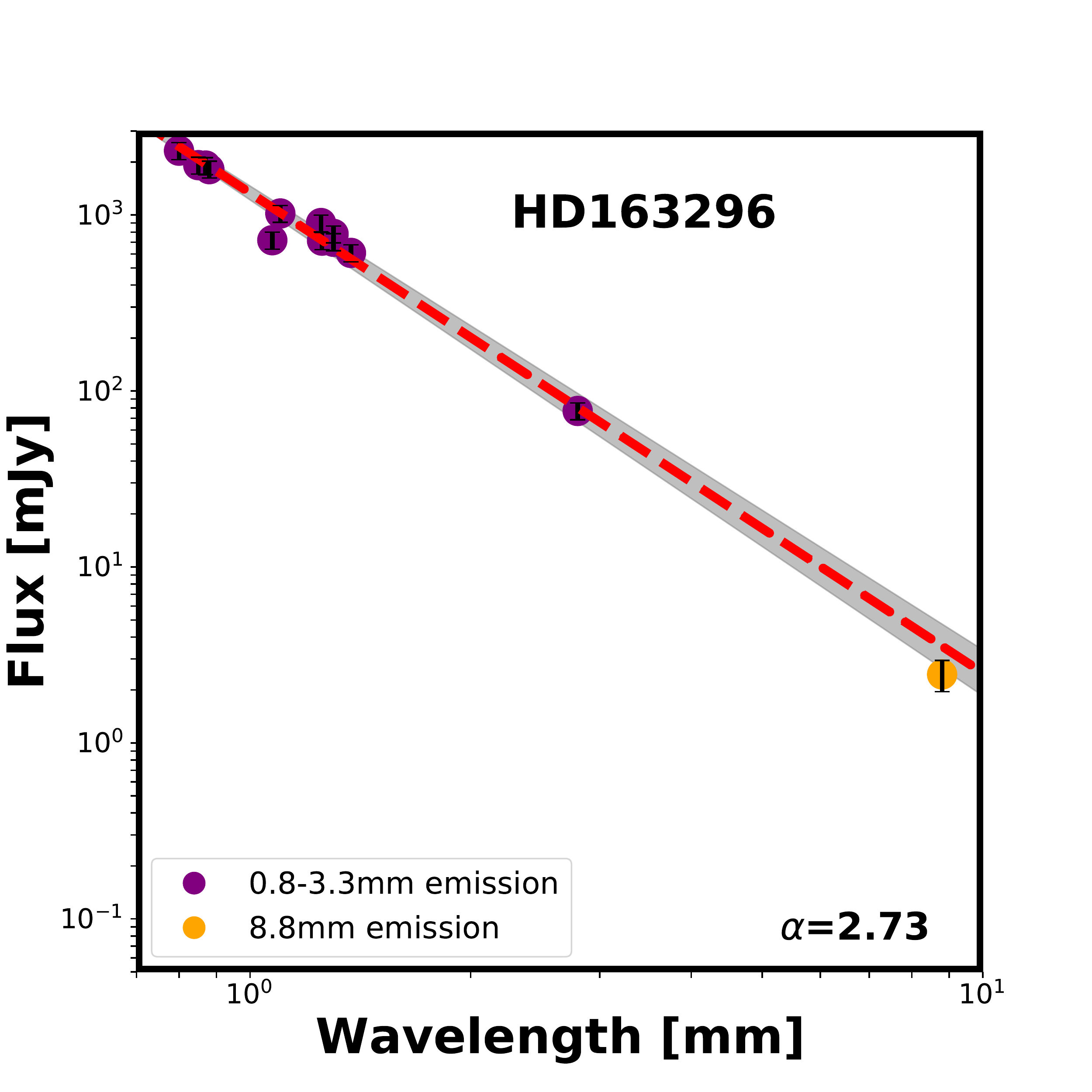}
	\caption{The spectral slope for HD163296. Purple markers represent \(\rm 0.8-3.3~mm\) observations from ALMA, and yellow markers represent 8.8~mm observations from ATCA. The error bars includes a calibration uncertainty of 10\% (ALMA) and 20\% (ATCA). Over-plotted with grey shading are linear least-squares fits error limits to parts of the spectral slope between 0.8 and 3.3~mm.}
	\label{fig:hd163296_sed}
\end{figure}

Here we present the 34~GHz observations of HD163296. This source was part of the Discs Down Under survey but is not classified as a transition disc. We include the continuum 34~GHz results in Table \ref{tab:hd163296} and show the continuum map in Figure \ref{fig:hd163296}. The continuum map shows significant central emission likely from the combination of both the inner disc and ring structure seen in the recent DSHARP survey \citep{2018ApJ...869L..49I}, our observations fail to resolve these features due to the lack of sufficiently long baselines and uv-coverage. No excess of emission is seen in Figure \ref{fig:hd163296_sed}.

\begin{table}
    \centering
    \caption{Integrated 7~mm flux, residual rms, synthesised beam size \(\theta_b\), and position angle for 34~GHz observations of HD163296.}
    \label{tab:hd163296}
    \begin{adjustbox}{width=0.48\textwidth}
        \begin{tabular}{lccccccc}
        \hline
        \multirow{2}{*}{Target} & \multirow{2}{*}{\begin{tabular}[c]{@{}c@{}}ATCA \\ 34~GHz flux \\ (mJy)\end{tabular}} &
        \multirow{2}{*}{\begin{tabular}[c]{@{}c@{}} ATCA \\ 1\(\rm\sigma\) rms \\  (mJy)\end{tabular}} & \multirow{2}{*}{\begin{tabular}[c]{@{}c@{}} ATCA \\ \(\rm \theta_{\rm beam}\) \\ (" x ")\end{tabular}} & \multirow{2}{*}{\begin{tabular}[c]{@{}c@{}} ATCA \\ \(\rm PA_{\rm beam}\) \\ (\(\degree\))\end{tabular}} \\ \\ \\ \midrule
        HD 163296               & 2.45 & 0.0026 & 1.31 x 0.400 & 4.59 \\ \bottomrule
\end{tabular}
    \end{adjustbox}
\end{table}

\section{\textsc{Frankenstein} hyper-parameter sweep}
\begin{figure*}
	\includegraphics[width=0.99\textwidth]{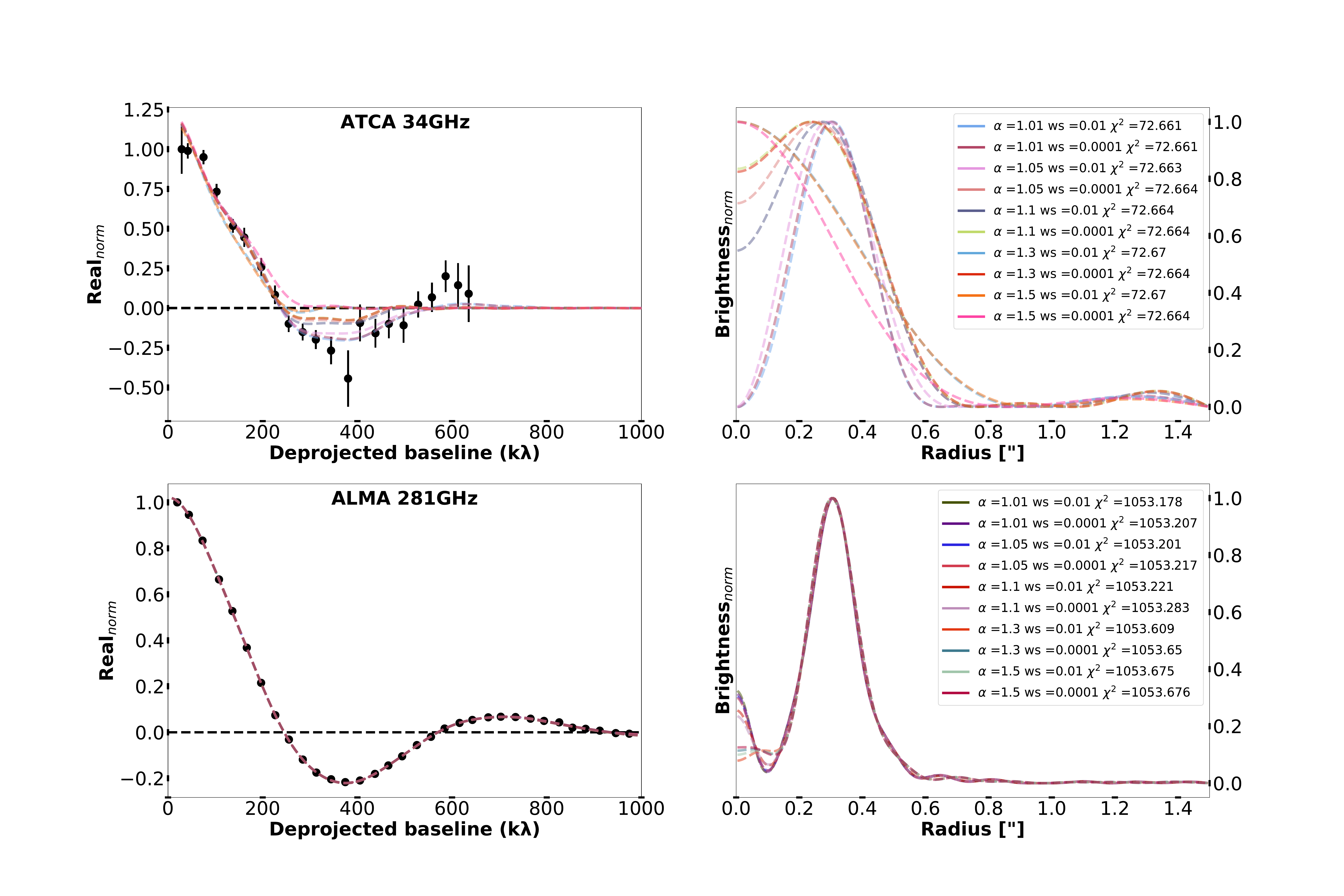}
	\caption{Frankenstein fit sensitivity to variations in both \(\rm \alpha\) and \(\rm w_{smooth}\), and the corresponding \(\chi^2\) values.}
	\label{fig:param_sweep}
\end{figure*}

Here we include the \textsc{Frankenstein} hyper-parameter sweep for HD100453. Figure \ref{fig:param_sweep} highlights the \textsc{Frankenstein's} fit dependence on the selection of \(\rm \alpha\) and \(\rm w_{smooth}\). The fit is not overly sensitive to \(\rm w_{smooth}\) which determines the length over which adjacent points are coupled. For our sample, \(\rm \alpha\) is varied between 1.01 and 1.5, the choice of \(\rm \alpha\) for each disc is determined by the characteristics of the visibility data. Values of \(\rm \alpha\) close to 1.01 are typically chosen for noisy data sets as \textsc{Frankenstein} attempts to fit longer baseline visibilities with lower SNR. As expected, Figure \ref{fig:param_sweep} shows that low choices of \(\rm \alpha\) result in lower \(\rm \chi^2\) values and \(\rm R_{peak}\) values that vary minimally. However, if \(\rm \alpha\) is sufficiently increased the data is fit poorly (as the SNR threshold is increased) and our Gaussian ring brightness profile model may become a Gaussian centred on \(r=0\). Given the relatively large \(\rm \chi^2\) values for these models, the difference in \(\rm R_{peak}\) values between \(r=0\) and \(r=R_{\rm ring}\) is not a concern.

\section{Varying the Phase Centre}
\begin{figure*}
	\includegraphics[width=0.99\textwidth]{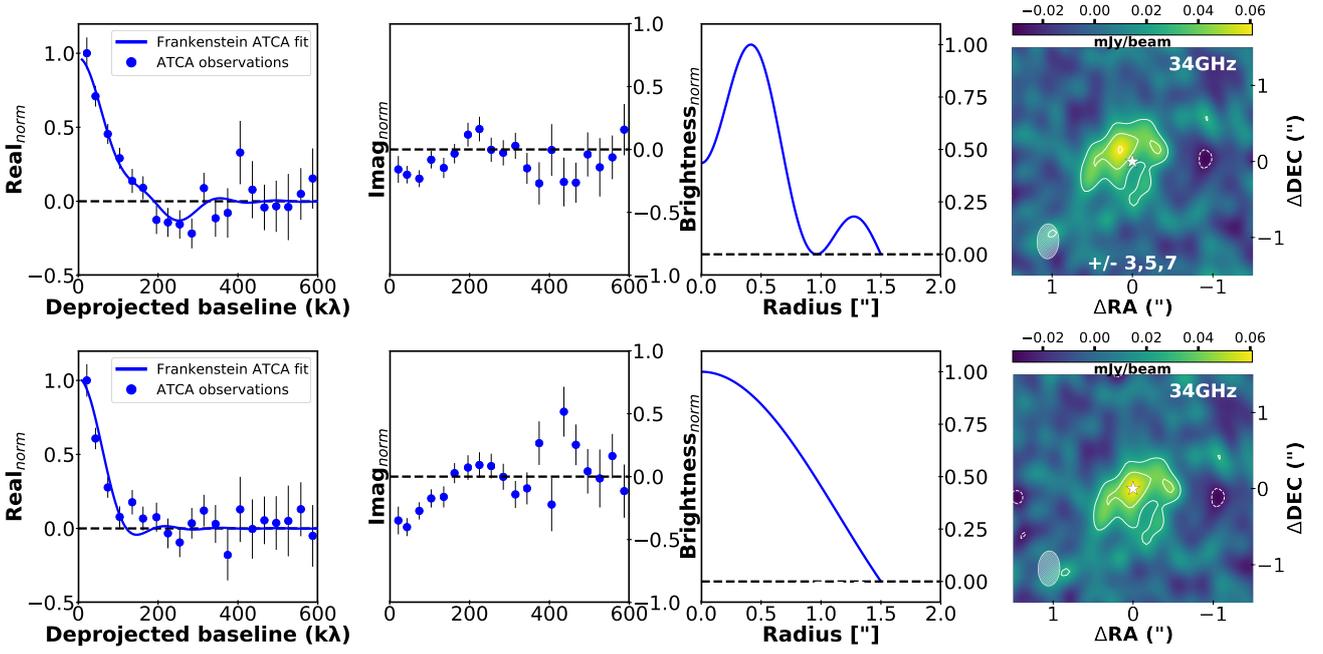}
	\caption{The effect of varying the phase centre of RY~Lup on the normalised real and imaginary components of the visibilities as a function of the deprojected baseline, normalised \textsc{Frankenstein} model radial brightness profiles, and the 8.8~mm ATCA continuum map at two different phase centres, with offsets $\rm dRA=0.15^{\prime\prime}$ and $\rm dDec=0.4^{\prime\prime}$.}
	\label{fig:ry_lup}
\end{figure*}

We present the visibility data and \textsc{Frankenstein's} fit dependence on the phase centre. For face-on sources with an obvious deficit of central emission and a clear ring-like morphology, phase centring is simple. However, for low SNR inclined/edge-on sources finding the exact phase centre can be troublesome. Figure \ref{fig:ry_lup} clearly indicates how changing the phase centre can either produce a Gaussian-like or Gaussian ring-like 1D radial brightness profile. It's inconsequential whether we to fit this non-parametrically with \textsc{Frankenstein} or with a parametric model, this is an intrinsic difficulty in fitting a radial profile to a non-axisymmetric disc with uncertain phase centre. Therefore, we suggest fitting these sources should be approached with caution.

\section{Imaginary Component of the Visibilities}
Here we include the normalised imaginary component of the visibilities as a function of the deprojected baseline for sources modelled with \textsc{Frankenstein} seen in Fig.~\ref{fig:vis}. For sources with relatively high SNR in the ATCA data (HD100453, SR24S, and HD169142) the imaginary component similarly has a high SNR and a low phase instability. For the rest of our sample, the imaginary component of the ATCA indicates large phase instabilities. 

\begin{figure*}
	\includegraphics[width=0.95\textwidth]{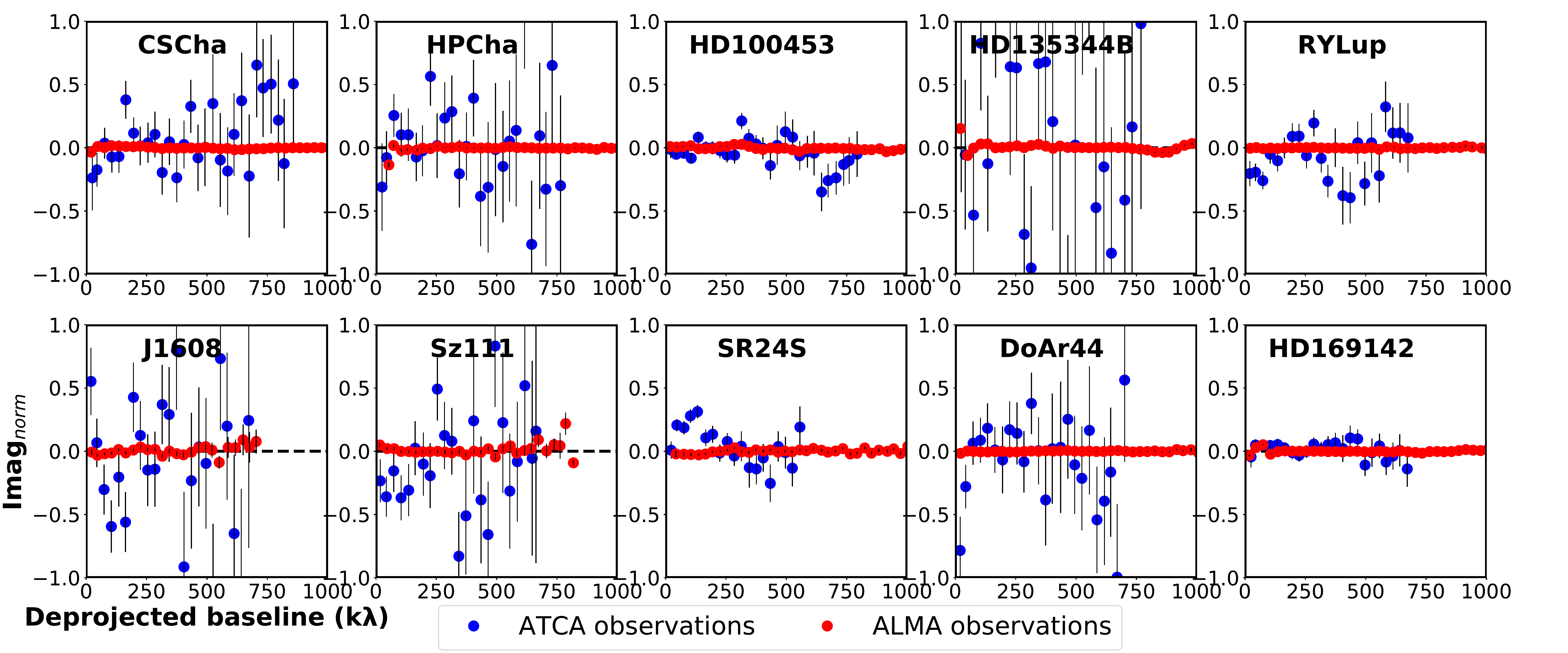}
	\caption{Normalised imaginary component of the visibilities as a function of the deprojected baseline for sources modelled with \textsc{Frankenstein} seen in Fig.~\ref{fig:vis}.}
	\label{fig:imag}
\end{figure*}

\section{Sources without Frankenstein ATCA Fits} \label{app:other_vis}
\begin{figure*}
	\includegraphics[width=0.9\textwidth]{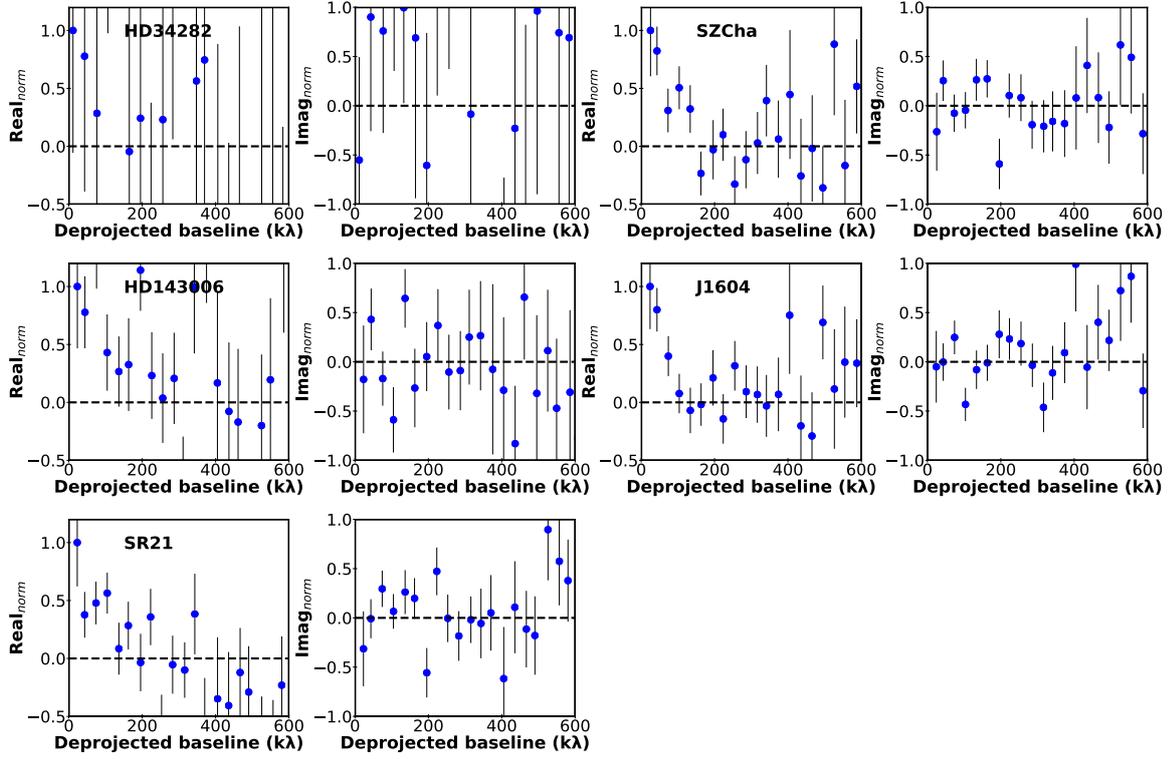}
	\caption{Normalised real and imaginary components of the visibilities as a function of the deprojected baseline for ATCA data with an SNR less than \(\rm 5\sigma\).}
	\label{fig:other_atca}
\end{figure*}

\begin{figure*}
	\includegraphics[width=0.9\textwidth]{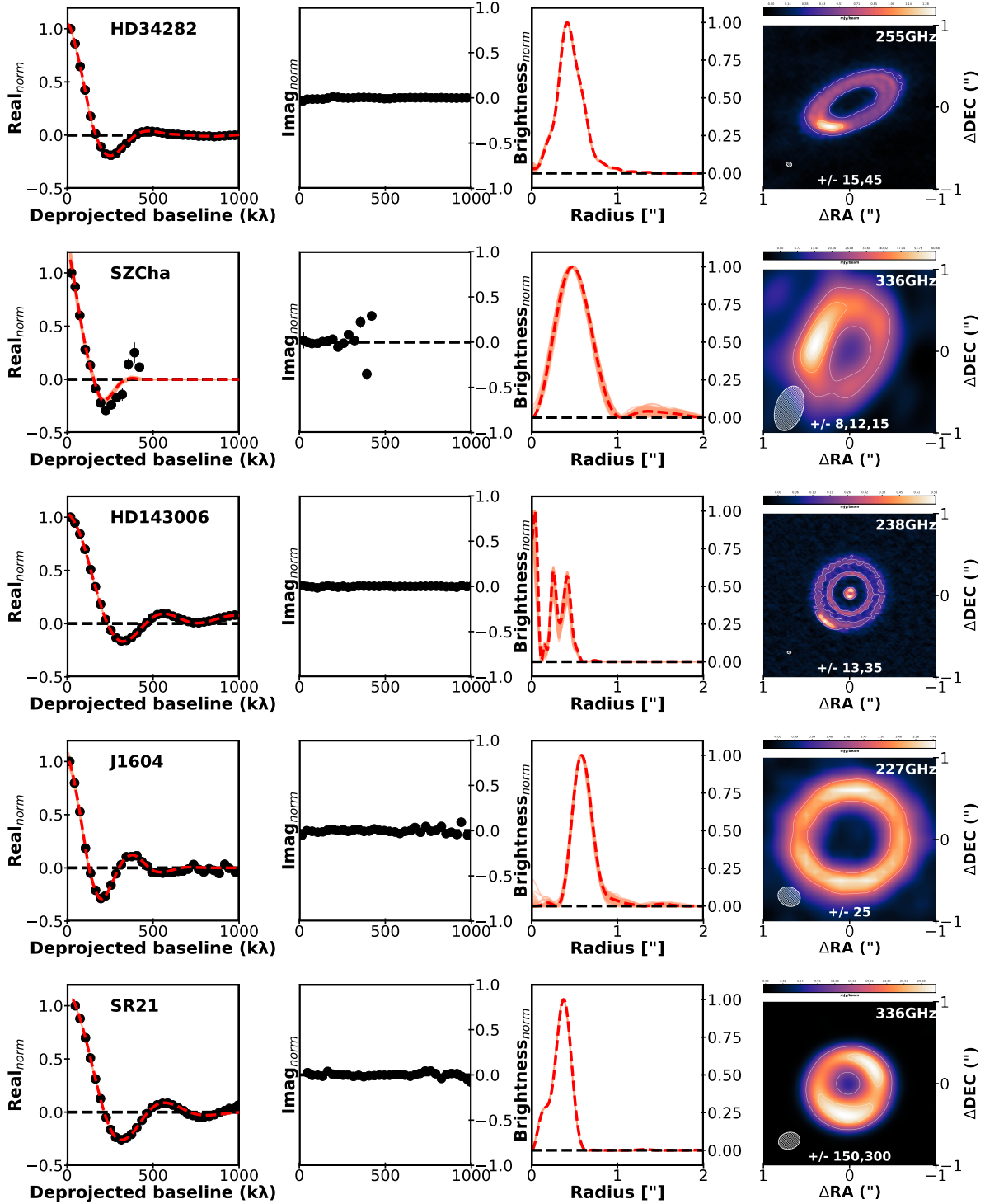}
	\caption{The normalised real and imaginary components of the visibilities as a function of the deprojected baseline, normalised \textsc{Frankenstein} model radial brightness profiles, and the sub-mm ALMA continuum map for sources not included in Figure \ref{fig:vis}.}
	\label{fig:other_alma}
\end{figure*}

We include the normalised imaginary component of the visibilities as a function of the deprojected baseline for ATCA data not modelled with \textsc{Frankenstein} and the corresponding ALMA data with \textsc{Frankenstein} models. We chose not to model the ATCA data of these sources due to the low SNR (\(\rm \leq 5\sigma\)) of the observations, this is obvious as shown by the large phase instability in Figure \ref{fig:other_atca}. In Figure \ref{fig:other_alma} the \textsc{Frankenstein} fits exhibit some deviation in the reconstructed 1D radial brightness profiles. This is likely due to asymmetrical structure seen in each of the corresponding continuum maps. In our analysis we use the same ALMA observations as \citet{2018ApJ...859...32P} for SZ~Cha (Band 7), J1604 (Band 6), and SR21 (Band 7). For HD34282 and HD143006, we make use of Band 6 data and \citet{2018ApJ...859...32P} analyses Band 7 data. For HD34282 our \(\rm R_{\rm peak}\) value is comparable to the radial location of the peak emission presented in \citet{2018ApJ...859...32P} (see Table \ref{tab:6}). However our models for SZ~Cha, HD143006, J1604, and SR21 deviate beyond the \(\rm 1\sigma\) error. The final calibrated visibilities for HD143006 was sourced from the DSHARP survey \citep{2018ApJ...869L..41A} and as a result, our analysis reveals more disc substructure in comparison to \citet{2018ApJ...859...32P}. For SZ~Cha, J1604, and SR21, our continuum maps show a peak emission at \(\rm r \approx 0.430, 0.594,\) and 0.355 arcseconds respectively, which is in close agreement with our \textsc{Frankenstein} fit \(\rm R_{peak}\) values. For SZ~Cha, we use \(\rm robust = 0 \) for our briggs weighting in CLEAN and shift our visibility data to the central deficit of emission by \(\rm dRA = -0.10\) and \(\rm dDec = -0.05\). The discrepancy in the model peak brightness between this work and \citet{2018ApJ...859...32P} for these three sources is likely due to limitations introduced by adopting a singular functional form of the disc structure to fit the visibilities parametrically. These limitations are most evident in the choice of the parametric model profile which will typically poorly fit long baseline data (\(\rm \geq 5 M\lambda\)). The accuracy of the fit to long baseline data (on any scale) strongly influences the recovered profile's features including the radial position of the peak brightness.

\begin{table}
 \centering
\caption{Radial brightness profile properties of the ALMA for sources not included in Figure \ref{fig:vis}. This includes the radial position of the peak brightness \(\rm R_{peak}\). Also included are the 1D profile peak positions from other models in the literature and the observed inclination and position angle of the target disc. Superscripts in the column headers refer to the position in the reference list of the last column.}
\label{tab:6}
\begin{adjustbox}{width=0.48\textwidth}
\begin{tabular}{@{}cccccc@{}}
\toprule
Source & \begin{tabular}[c]{@{}c@{}}$R_{\rm peak}$\\ obs\\ (")\end{tabular} & \begin{tabular}[c]{@{}c@{}}$R_{\rm peak}$\\ \(\rm Lit^{a}\) \\ (")\end{tabular} & \begin{tabular}[c]{@{}c@{}}\(\rm Incl.^{b}\) (\(\degree\)) \end{tabular}  & \begin{tabular}[c]{@{}c@{}}\(\rm PA^{d}\) (\(\degree\))  \end{tabular} & \begin{tabular}[c]{@{}c@{}}Ref. \\ (a,b) \end{tabular} \\ \midrule
HD34282 & \(0.416_{-0.001}^{+0.001}\) & \(0.427_{-0.001}^{+0.001}\) & 59 & 117 & 1,2 \\ \\
SZ~Cha & \(0.476_{-0.015}^{+0.015}\) & \(0.375_{-0.011}^{+0.011}\) & 47 & 154 & 1,1 \\ \\
HD143006 & \(0.039_{-0.014}^{+0.014}\) & \(0.506_{-0.006}^{+0.006}\) & 29.5 & 151 & 1,1 \\ \\
 & \(0.254_{-0.001}^{+0.001}\)* & & & & \\ \\
 & \(0.413_{-0.001}^{+0.001}\)** & & & & \\ \\
J1604 & \(0.586_{-0.007}^{+0.007}\) &  \(0.553_{-0.010}^{+0.010}\) & 6 & 80 & 1,2 \\ \\
SR21 & \(0.377_{-0.003}^{+0.003}\) & \(0.424_{-0.001}^{+0.001}\) & 16 & 14 & 1,2 \\\bottomrule
\end{tabular}
\end{adjustbox}
    {\centering \textbf{Notes:} *Second Peak \\ **Third Peak \\ \textbf{References} (1) \citet{2018ApJ...859...32P}, (2) \citet{2020francis} \par}
\end{table}

\bsp	
\label{lastpage}
\end{document}